\documentstyle[12pt,aasms4,epsf]{article}

\lefthead{C.~Alcock et al.}
\righthead{MACHO observations of LMC Type~{\large\bf II} Cepheids and 
RV~Tauri stars}

\begin{document}

\title {The MACHO project LMC variable star inventory: The discovery of 
RV~Tauri stars and new Type~{\Large\bf II} Cepheids in the LMC}

\author{C.~Alcock,\altaffilmark{1,2} R.A.~Allsman,\altaffilmark{3} 
D.R.~Alves,\altaffilmark{1,4} T.S.~Axelrod,\altaffilmark{5} 
A.~Becker,\altaffilmark{2,6} 
D.P.~Bennett,\altaffilmark{1,2,7} K.H.~Cook,\altaffilmark{1,2} 
K.C.~Freeman,\altaffilmark{5} 
K.~Griest,\altaffilmark{2,8} J.A.~Guern,\altaffilmark{2,8}
W.A.~Lawson,\altaffilmark{9} M.J.~Lehner,\altaffilmark{2,8} 
S.L.~Marshall,\altaffilmark{1,2} 
D.~Minniti,\altaffilmark{2} 
B.A.~Peterson,\altaffilmark{5}
Karen R. Pollard,\altaffilmark{10} M.R.~Pratt,\altaffilmark{2,6} 
P.J.~Quinn,\altaffilmark{11}
A.W.~Rodgers,\altaffilmark{5}
C.W.~Stubbs,\altaffilmark{2,6} and W.~Sutherland\altaffilmark{12}
}
\altaffiltext{1}{Lawrence Livermore National Laboratory, Livermore, 
CA94550, U.S.A.}
\altaffiltext{2}{Center for Particle Astrophysics, University of 
California, Berkeley, CA94720, U.S.A.}
\altaffiltext{3}{Supercomputing Facility, Australian National 
University, Canberra, ACT 0200, Australia}
\altaffiltext{4}{Dept. of Physics, University of California, 
Davis, CA95616, U.S.A.}
\altaffiltext{5}{Mt Stromlo and Siding Spring Observatories, 
Australian National University, Weston, ACT 2611, Australia}
\altaffiltext{6}{Dept. of Astronomy and Physics, University of 
Washington, Seattle, WA98195, U.S.A.}
\altaffiltext{7}{Dept. of Physics, University of Notre Dame, 
Notre Dame, IN46556, U.S.A.}
\altaffiltext{8}{Dept. of Physics, University of California, 
San Diego, La Jolla, CA92093, U.S.A.}
\altaffiltext{9}{Dept. of Physics, Australian Defence Force 
Academy, University of New South Wales, Canberra, ACT 2600, Australia}
\altaffiltext{10}{South African Astronomical Observatory, PO Box 9, 
Observatory 7935, South Africa}
\altaffiltext{11}{European Southern Observatory, Karl-Schwarzchild 
Str. 2, D-85748, Garching, Germany}
\altaffiltext{12}{Dept. of Physics, University of Oxford, Oxford, 
OX1 3RH, U.K.}

\begin{abstract} 
We report the discovery of RV~Tauri stars in the Large Magellanic Cloud. In
light and colour curve behaviour, the RV~Tauri stars appear to be a direct
extension of the Type~{\small II} Cepheids to longer periods.  A single
period--luminosity--colour relationship is seen to describe both the
Type~{\small II} Cepheids and the RV~Tauri stars in the LMC. We derive the
relation: $V_{\circ} = 17.89 (\pm 0.20) - 2.95 (\pm 0.12) \log_{10}P + 5.49
(\pm 0.35) \overline{(V-R)_{\circ}}$, valid for Type~{\small II} Cepheids
and RV~Tauri stars in the period range $0.9 < \log_{10}P < 1.75$. Assuming a
distance modulus to the Large Magellanic Cloud of 18.5, the relation in
terms of the absolute luminosities becomes: $M_{V} = -0.61 (\pm 0.20) - 2.95
(\pm 0.12) \log_{10}P + 5.49 (\pm 0.35) \overline{(V-R)_{\circ}}$.
\end{abstract}

\section{Introduction}

Pulsating variables which occupy the population~{\small II} instability
strip (for example RV~Tauri, W~Virginis, BL~Herculis and RR~Lyrae stars)
represent an important testing ground for our theories of stellar evolution
and pulsation. As low-mass stars of the intermediate disk and halo
population they are useful probes of our Galaxy's halo and bulge as well as
the outer regions of nearby galaxies.
 
The period boundaries for Type~{\small II} Cepheids are determined by the
transition from the RR~Lyr stars at the short period end to the RV~Tauri and
long period variables at the long period limit (for a comprehensive review
of the population~{\small II} Cepheids, see Wallerstein \& Cox 1984).  Those
stars with periods between about 1 and 8 days are usually designated BL~Her
variables while the longer period variables are known as W~Vir stars. This
nomenclature is useful because the evolutionary state of the two groups is
different. The BL~Her variables are stars evolving through the instability
strip (IS) from the Horizontal Branch towards the asymptotic giant branch
(AGB) following the exhaustion of helium in their cores. The W~Vir variables
are believed to be hydrogen- and helium-shell burning stars which are making
blue-loop excursions into the IS from the AGB.

The RV~Tauri stars are semiregular pulsating variables which are located in
the brightest part of the population~{\small II} instability strip.  As
such, there is some overlap in photometric properties with the W~Vir 
Type~{\small II} Cepheids.  A defining characteristic of the RV~Tauri
variables is a light curve which displays alternating deep and shallow
minima. The period between adjacent deep minima is generally in the range
40--150 days and these stars have spectral types F--K with luminosity class
Ia--II. Unlike the W~Vir stars, most field members of the RV~Tauri
class exhibit strong infrared excesses indicative of extensive amounts of
circumstellar material. From an evolutionary perspective they are thought 
to be low-mass objects at the termination of their AGB evolution.

Of fundamental importance to our understanding of the RV~Tauri variables,
and their relationship to the Type~{\small II} Cepheids, is a knowledge of
their physical properties. In particular, our understanding of their mass
and evolutionary state depends critically on their luminosities. However,
these luminosities are poorly known. There are currently no published
identifications of RV~Tauri stars in the Magellanic Clouds. Furthermore,
there is concern that the handful of stars which are globular cluster
members (and for which we have some knowledge of their luminosities) are not
of the same spectroscopic type or metallicity as the field RV~Tauri stars.

No definitive period--luminosity relation exists. The most commonly used
relation (DuPuy 1973) is based on observations of a very small number of
low-metallicity globular cluster members and little agreement is seen
between this relation and luminosities derived from recent spectroscopic
observations of field RV~Tauri stars (Wahlgren 1992). Period--luminosity
relations exist for the Type~{\small II} Cepheids (Demers \& Harris 1974; Harris
1985; McNamara \& Pyne 1994; Nemec, Nemec \& Lutz 1994) but in general,
these are valid only for shorter periods and the RV~Tauri variables are
often explicitly excluded from the analyses. Nemec et al.\ (1994) revived an
interesting hypothesis (Arp 1955) that the P-L relation for globular cluster
Type~{\small II} Cepheids (which also included some cluster RV~Tauri stars)
was actually two parallel relations defined by a harmonic pulsation mode
sequence and a first-overtone pulsation mode sequence.\\

In consequence, the primary goals of this study were as follows:
\begin{enumerate}
\item to discover whether variables of the RV~Tauri type do exist in the
LMC, and
\item to investigate the relationship between the 
Type~{\small II} Cepheids and the RV~Tauri stars, in particular to determine 
whether there is a common period--luminosity relationship.
\end{enumerate}

The MACHO project photometry of Large Magellanic Cloud (LMC) variables
provides us with an ideal database address the above two issues since we
are able to observe both classes of variables in a common environment. The
advantages of this are that all stars are at a known common distance,
differential reddening is low and large numbers of observations on a common
photometric system are available over many cycles. When only smaller or more
fragmentary light curve datasets are available, the RV~Tauri stars can
easily be confused with semi-regular or eclipsing variables.

\section{Previous observations of Type~{II} Cepheids in the LMC}

The primary sources of optical observations of Type~{\small II} Cepheids in
the LMC are those of Gaposchkin (1970) and Payne-Gaposchkin (1971). These
researchers produced photographic light curves, variable type
classifications and determined periods for LMC variables from the analysis
of Harvard photographic plates. A total of seventeen Type~{\small II}
Cepheids in the LMC were listed in Payne-Gaposchkin (1971). Although some of
these Type~{\small II} Cepheids have periods as long as RV~Tauri stars, none
were classified as such from their light curves. Harris (1985) states that
there are 20 known Type~{\small II} Cepheids in the LMC, but one (HV13064)
has since been shown to be a highly-reddened Type~{\small I} Cepheid using
infrared observations (Laney 1991). 

Infrared studies of LMC Type~{\small II} Cepheids include those by Welch
(1987) and Laney (1991). Welch (1987) obtained single-phase JHK observations
of nine LMC Type~{\small II} Cepheids, three of which show infrared excesses
at K.  Welch suggested that these K-excess stars are simply related to
RV~Tauri variables. Laney (1991) obtained several JHK observations each of
nineteen LMC Type~{\small II} Cepheid candidates, confirming that four had
infrared excesses in the K band.  

\section{Observations}

The MACHO project has been described by Alcock et al.\ (1992) and Alcock et
al.\ (1995). A dedicated 1.27-m telescope at Mount Stromlo, Australia is
used to obtain observations of the LMC year-round (Hart et al.\ 1996).  The
camera built specifically for this projects (Stubbs et al.\ 1993) has a
field of view of 0.5 square degrees, which is achieved by imaging at prime
focus. Photometric observations of the LMC fields are obtained in two
bandpasses simultaneously, using a dichroic beamsplitter to direct the
`blue' (440--590\,nm) and `red' (590--780\,nm) light onto 2$\times$2 mosaics
of 2048$\times$2048 Loral CCDs. (Hereafter we refer to these bandpasses as
$V_{\rm M}$ and $R_{\rm M}$ respectively.) The 15$\mu$m pixels map to 0.63
arcsec on the sky.  The data were reduced using a profile-fitting photometry
routine known as {\sc SODOPHOT}, derived from {\sc DoPHOT} (Mateo \&
Schechter 1989). The output photometry contains flags indicating suspicion
of errors due to crowding, poor seeing, array defects and radiation events.
For more details of observations and photometric reductions see Alcock et
al.\ 1996a.

For the present study, use was made of the first four years of LMC data,
consisting of some 20000 frames distributed over 22 fields concentrated
along the bar of the LMC. This sample contains a total of approximately 8
million stars, of which over 40\,000 have been found to be variable stars 
-- most newly discovered. Typically, the dataset for a given star covers a
timespan of about 1500 days and usually contains 300--1100 good photometric
measurements.

\section{Analysis}

A number of selection criteria were applied to the database of MACHO project
variable star photometry in order to produce a list of possible Type~{\small
II} Cepheid and RV~Tauri star candidates. These basic criteria were:
\begin{enumerate}
\item 8 d $<$ Period $<$ 100 d \\
\item $18.5 - 3 \log P$ $< R <$ 18.0 \\
\item $0.3 < (V-R) < 0.6$.
\end{enumerate}

Although Type~{\small II} Cepheids with periods shorter than 8\,d are
present in the MACHO project database, these data were not available for
this project. The brighter limit on the $R$ magnitude was chosen so as to
exclude the Type~{\small I} Cepheids. The selection in terms of $(V-R)$
colour is representative of colours of galactic Type~{\small II} Cepheids
and RV~Tauri stars. It is quite possible that some relevant variables may
have been omitted by the choice of criteria above and it is hoped that this
study will be extended in the future by relaxing some of these criteria. In
particular it would be useful to extend the long period limit and also to
choose a broader range of colour, specifically to include variables with
larger values of $(V-R)$.

The $V_{\rm M}$ and $R_{\rm M}$ light curves of the resulting selection of
approximately 250 variables were searched for periodicities using a standard
fourier-type period-finding code. The phased $V_{\rm M}$ light and
$(V-R)_{\rm M}$ colour curves were then visually examined in order to
classify the variables.  RV~Tauri light curves can often be confused with
eclipsing binary light variations making the colour information very
important in this classification process.

\section{Results}
\subsection{Lightcurves}

Thirty-three Type~{\small II} Cepheid and RV~Tauri candidates were discovered.
Finder charts for these variables are shown in Figure~\ref{figfinder}. The
internal star identifier, the JD2000.0 coordinates and the fundamental
period found from the period analysis are listed in Table~\ref{tabdata1}. A
cross check of these variables with the list of seventeen LMC 
Type~{\small II} Cepheids given in Payne-Gaposchkin (1971) revealed six stars 
in common. These designations are included in Table~\ref{tabdata1}. Two stars
designated as Type~{\small II} Cepheids in Payne-Gaposchkin (1971) are also
in the MACHO database but we were unable to confirm the Type~{\small II}
Cepheids or RV~Tauri star classification due to lack of photometry. These
stars (HV~2522 = MACHO*05:26:27.2-66:42:58 and HV~2351 =
MACHO*05:16:35.6-68:21:02) were therefore not included in any further
analysis but will continue to be observed. The remaining nine HV stars were
either not in our MACHO fields, were in a MACHO field without a template or
were unlocated due to position errors.

We have identified five of the six Harvard variables listed in
Table~\ref{tabdata1} as RV~Tauri candidates rather than Type~{\small II}
Cepheids.  Indeed, Gaposchkin (1970) comments on the fact that the majority
of the Type~{\small II} Cepheids discovered in the Harvard survey have long
periods: \begin{quote} ``$\ldots$ only four of them have periods that are
close to that of the prototype W Virginis. The other 12 are clustered around
a period more than twice as large.'' \end{quote}
 
Phased $V_{\rm M}$ light and $(V-R)_{\rm M}$ colour curves for all the new
Type~{\small II} Cepheid and RV~Tauri variables discovered in the MACHO
database to date are shown in Figures~\ref{figlightcurves} and
\ref{figcolourcurves}.  The photometric data have been phased using the
fundamental period for the Type~{\small II} Cepheids and the formal period
for stars with periods greater than 20 days.  An approximate epoch was
chosen such that the `primary' (or faintest) light minimum of the star
occurs at phase 0.0.  Only data free from suspected error are plotted.
Typical photometric uncertainties are in the range 1.5--2 per cent. The 
light curves
were qualitatively classified as sinusoidal, crested, flat-topped
(following the scheme proposed by Kwee 1967), and RV~Tauri-like and these
designations are included in Table~\ref{tabdata1}.

\subsection{Fourier fits and decomposition parameters}

The phased $V_{\rm M}$ light curve of each Type~{\small II} Cepheid or
RV~Tauri candidate was fitted with a truncated Fourier series of the
fundamental period and up to the 9th order. Typically a fifth order fourier
fit was found to be a good representation of the data. Figure~\ref{figffit}
shows the 5th order fourier fit to the MACHO photometric data for 
MACHO*05:37:45.0-69:54:16 which has a double or `formal' period of
60.816\,d. This star displays consistent deep--shallow alternations of its
light curve minima and some variability in the depths of the minima is
apparent. 

The fourier decomposition
parameters $\Phi_{21}$, $R_{21}$, $\Phi_{31}$, and $R_{31}$ are plotted
versus $\log P$ in Figure~\ref{figfourier}. For stars with periods longer
than 20\,d, the fourier fit using the fundamental period is an increasing
poor fit as the light curve becomes more RV~Tauri-like. The fourier
decomposition parameters become increasingly uncertain as the light curves
becomes dominated by the double periodicity.
Figure~\ref{figfourier} shows there is a progression in $\Phi_{31}$ which
increases for increasing $\log P$ for all periods. 

\subsection{Conversion to standard system and absolute luminosities}

The $V_{\rm M}$ and $R_{\rm M}$ bandpasses were converted to
Kron-Cousins (KC) $V$ and $R$ bandpasses using the latest transformations
determined from the ongoing internal calibrations of the MACHO database.
Mean $V$ and $R$ magnitudes and mean $(V-R)_{\rm KC}$ colours were
calculated for each star. To determine the dereddened mean magnitudes we
initially adopted an $\overline{E(B-V)}$ to the LMC of 0.074 (Caldwell \&
Coulson 1985). The standard value of the total to selective extinction,
$R_{V}$ = $A_{V}/E(B-V) = 3.1$, was used (Cousins 1980). Using the
normalised interstellar reddening curve (Whitford 1958) we derive
$A_{R}/A_{V} = 0.79$ and hence $\overline{E(V-R)} = 0.048$.

Dereddened mean magnitudes were also determined by adopting individual
foreground reddenings from the map of galactic foreground colour excess
towards the LMC published by Schwering \& Israel (1991). These $E(B-V)$
values ranged between 0.07 and 0.15 for the stars in our sample. It was
found that adopting these individual foreground reddenings had no
significant effect on the derived period--luminosity relationship within the
stated uncertainties.

In both the calculations outlined above, no contribution to the reddening
from circumstellar material was included. Although it is unlikely that there
is much, if any, circumstellar reddening for the Type~{\small II} Cepheids
it it possible that some circumstellar reddening exists for the RV~Tauri
stars. One star in our sample (MACHO*05:14:18.1-69:12:35) is known to
possess an infrared excess at K (Laney 1991). However, as we have no
independent means of determining the individual circumstellar reddenings for
the other stars in the sample we have ignored this possible contribution.

A distance modulus of 18.5 to the LMC was used to derive the absolute
luminosities of the Type~{\small II} Cepheids and RV~Tauri stars. 
Table~\ref{tabdata2} presents the mean intrinsic magnitude ($V$), the
adopted $E(B-V)$, the mean dereddened magnitudes ($V_{\circ}$ and
$R_{\circ}$), the mean intrinsic colour ($(V-R)_{\circ}$), the quantity
$W_{V}$ ($=V_{\circ}-5.49(V-R)_{\circ}$) and the mean absolute luminosity
($M_{V}$). The dereddened HR diagram for the MACHO Type~{\small II} Cepheids
and RV~Tauri stars is shown in Figure~\ref{fighr}. For the same intrinsic
colour the Type~{\small II} Cepheids have lower luminosities than
the RV~Tauri stars. This conclusion will only be strengthened if the
RV~Tauri stars possess some degree of circumstellar extinction.
It is interesting to note that the two
Type~{\small II} Cepheids which fall close to the RV~Tauri star `locus' 
are stars with periods around 20\,d which are beginning to show 
some RV~Tauri characteristics.

\section{Discussion}
\label{sectdiscuss}

In general it was found that Type~{\small II} Cepheids with periods greater
than $\sim$20 days display increasing variability in the depth, shape and
phasing of their light curve minima.  The periodograms of these stars also
exhibit increasing strength in the subharmonic frequency (or the double
period). When plotted on this double or formal period, as in
Figure~\ref{figlightcurves} and \ref{figcolourcurves}, the different
behaviour of the two minima (and maxima) becomes apparent.  Classic RV~Tauri
behaviour (Pollard et al.\ 1996) is displayed by many of the stars with
`single' or fundamental periods greater than about 20 days, that is:
\begin{itemize}
\item alternating deep (`primary') and shallow (`secondary') minima;
\item secondary minima (and maxima) more variable than primary minima 
(and maxima);
\item bluest colours during the rising branch of the light curve, particularly
during the rise from the secondary minima.
\end{itemize}

Figure~\ref{figperlac} plots the luminosity, amplitude and colour on the
MACHO photometric system versus period.  Different symbols are used for the
different light curve types in order to reveal any trends.  This figure
shows a gradual transition in behaviour from the short period Type~{\small
II} Cepheids to the longer period RV~Tauri stars. The $V_{\rm M}$ magnitude
versus $\log P$ plot (Figure~\ref{figperlac}(a)) displays a gradually
increasing slope for the longer period stars.  The amplitude of variation
increases (Figure~\ref{figperlac} (b)) for increasing period up to a maximum
amplitude at around 20\,d and then decreases for longer period stars. 
Results from recent theoretical studies find increasing complexity and lower
amplitudes for the more luminous, low-mass stars of intermediate $T_{\rm
eff}$ due to strong non-adiabatic and non-linear effects.  The observed
preference towards lower amplitudes for higher luminosities noted in
Figure~\ref{figperlac} is consistent with what is seen for post-AGB
transition objects (Sasselov 1992).

As found by previous researchers (Harris 1985, Nemec et al.\ 1993), the
($V-R$) colours are seen to get redder for longer period Type~{\small II}
Cepheids. An interesting trend seen in Figure~\ref{figperlac} (c) is that
the ($V-R$) colours get bluer again for the RV~Tauri stars of longer period.
This behaviour has not been noted previously for the galactic RV~Tauri
stars. However it should be noted that stars with colours redder than
($V-R$)\,=\,0.6 were excluded using our selection criteria and it is
possible that this may have some effect on this colour trend.

Two stars (MACHO*05:32:54.5-69:35:13 and MACHO*05:40:00.5-69:42:14) are
unusual and their classification is tentative. The first star has a very low
amplitude light curve with slight variability in the depth of its minima but
the period is relatively stable. It appears more like a long-period,
low-amplitude W~Vir star than an RV~Tauri star. The second star displays
minima of variable amplitude, but not consistently alternating, which is
more reminiscent of some irregular RV~Tauri stars or some semiregular
variables.  As the classification of these two stars is only tentative at
this stage, they were not included in the following analysis of the
period--luminosity relationship.

\subsection{Period--Luminosity relationship}

The MACHO database of LMC photometry allows us a direct interpretation of
the luminosity and hence the P--L and P--L--C relations for these variables.
We have examined the P--L and the P--L--C relations of the Type~{\small II}
Cepheids and RV~Tauri stars in the LMC using both the MACHO light curves and
the $V$ and $R$ light curves in the standard (Kron-Cousins) system.  Three
stars in Table~\ref{tabdata1} were not included in this
analysis. These stars were the two stars of tentative classification
described above. The other star excluded from the analysis was HV~5756
(MACHO*05:19:26.9-69:51:52), an eclipsing binary Type~{\small II} Cepheid
(Alcock et al.\ 1996b) which appears to be brighter and bluer than if it were
a single star.

In Figure~\ref{figperlumold} (a) we plot the galactic P--L relation for
Type~{\small II} Cepheids taken from a number of sources (see also
Table~\ref{tabpl}). The absolute magnitudes of the LMC Type~{\small II}
Cepheids and RV~Tauri stars derived in this study are also plotted in this
figure. The relations derived from galactic Type~{\small II} Cepheids
appear to be a reasonable representation of the P--L relation for the LMC
Type~{\small II} Cepheids but the extrapolations of these relations to the
longer period RV~Tauri stars is problematic. The RV~Tauri stars appear to be
more luminous that the extrapolated P--L relations would suggest. The
criteria for classifying the Type~{\small II} Cepheids as fundamental or
harmonic pulsators by Nemec et al.\ (1994) are unclear and we have been
unable to apply these designations to the LMC Type~{\small II} Cepheids and
RV~Tauri stars.  The relation of DuPuy (1973) is clearly anomalous, but this
is probably due to the fact that it based on observations of only three
confirmed globular cluster RV~Tauri stars.

We have derived a single period--luminosity relationship for the
Type~{\small II} Cepheids and RV~Tauri stars in the LMC:
\begin{equation}
\begin{array}{rll}
M_{V} = & 1.34 \;\; - & 3.07 \log P \\
        &\pm0.45      &\pm 0.35      \\
\end{array}
~\hspace{5mm} \sigma = 0.44    
\end{equation}
valid for $0.9 < \log P < 1.75$. This P--L relation is displayed in
Figure~\ref{figperlumold} (b). This equation is quite similar to the
relation for longer period Type~{\small II} stars derived by Harris (1985),
but there seems to be a alight systematic deviation from the derived
relation by the shortest period stars. The possibility that a colour effect
is present was investigated by including a colour term in the multivariate
regression analysis. The main contributor to the colour term is the
differing effective temperatures of the stars. It also possible that there
is some difference in the extinction towards the stars in the sample.

A multivariate linear regression in $\log P$ and $(V-R)_{\circ}$ with
$V_{\circ}$ as the independent variable allows us to derive a P--L--C
relation for the 30 LMC Type~{\small II} Cepheids and RV~Tauri stars:
\begin{equation}
\begin{array}{rlll}
V_{\circ} = & 17.89 \;\; - & 2.95\log P \;\; + & 5.49\overline{(V-R)_{\circ}}\\
            & \pm0.20      & \pm0.12           & \pm0.37\\
\end{array}
~\hspace{5mm} \sigma = 0.15    
\end{equation}
valid for $0.9 < \log P < 1.75$. 
If $\log P$ is adopted as the independent variable, then the inverse
P--L--C relation becomes:  
\begin{equation}
\begin{array}{rlll}
V_{\circ} = & 18.06 \;\; - & 3.08\log P \;\; + & 5.47\overline{(V-R)_{\circ}}\\
            & \pm0.19      & \pm0.13           & \pm0.45\\
\end{array}
~\hspace{5mm} \sigma = 0.15    
\end{equation}
valid for $0.9 < \log P < 1.75$. 
In terms of absolute magnitudes, and adopting a distance modulus to the LMC
of 18.5, the direct P--L--C relation becomes:
\begin{equation}
\begin{array}{rlll}
M_{V} = & -0.61 \;\; - & 2.95 \log P \;\; + & 5.49 \overline{(V-R)_{\circ}} \\
        & \pm0.20      & \pm 0.12           & \pm 0.37 \\
\end{array}   
\end{equation}
The quantity $W_{V}$ ($=V_{\circ}-\alpha(V-R)_{\circ}$) is thus a projection
of the P--L--C relation which removes the largest part of the effect of
differing effective temperatures and differential absorption (Alcock et al.\
1995). The plot of $W_{V}$ ($=V_{\circ}-5.49(V-R)_{\circ}$) versus 
$\log P$ is shown in Figure~\ref{figpl}. This figure indicates that the
RV~Tauri stars are a direct extension of the Type~{\small II} Cepheids to
longer periods and that these population~{\small II} variables are more
accurately represented by a P--L--C relation than a P--L relation (the
scatter about the regression line is less than half that shown in
Figure~\ref{figperlumold} where the colour term is omitted).

All the candidate RV~Tauri stars that we have identified from the MACHO
project LMC variable star database are very luminous variables. This result
lends very strong support to the hypothesis that these stars are in the
post-AGB stage of their evolution.  The RV~Tauri stars therefore appear to
represent an important probe of this critical phase of late stellar
evolution. The strong link between the behaviour of the Type~{\small II}
Cepheids and the RV~Tauri stars is further evidence for a low-mass
interpretation for the RV~Tauri variables and also may suggest an
evolutionary connection between these two classes of variables.

\section{Conclusions}

A study of the MACHO project LMC variable star database has shown conclusive
evidence that variable stars of the RV~Tauri class do exist in the LMC.
These stars show light and colour curve characteristics that are very
similar to their galactic counterparts. A total of thirty three Type~{\small
II} Cepheids and RV~Tauri stars candidate were identified in the LMC,
although two of these have only tentative classifications. An additional two
stars identified as Type~{\small II} Cepheids in Payne-Gaposchkin (1971)
have too little MACHO photometry to confirm their classification.

The discovery of RV~Tauri stars in the LMC has enabled us to finally derive
accurate absolute luminosities for these objects.  The derived absolute
magnitudes of the new LMC RV~Tauri variables of about --4.5 for variables
with fundamental periods of about 50\,d ($\log P = 1.7$) lends strong
support to the proposal that these objects are luminous, post-AGB stars
evolving to the left in the HR diagram.

A progression in behaviour from the short period Type~{\small II} Cepheids
through to the longer period RV~Tauri stars in terms of their light and
colour curve properties was noted. The variables with periods greater than
about 20\,d show increasingly strong RV~Tauri characteristics.  

A single period--luminosity--colour relation is seen to describe both the
Type~{\small II} Cepheids and the RV~Tauri stars in the LMC. The continuity 
in behaviour and properties is strong evidence for an evolutionary
connection between these two classes of variables.

\section{Future Work}

There are a a number of refinements and extensions to the above study
that can be made. To discover additional RV~Tauri stars in the LMC we need
to relax our selection criteria on the MACHO variable star database in order
to include variables with longer periods and redder colours. This relaxation
of the selection criteria should also allow us to investigate whether any
bias was introduced in our derived P--L--C relation. The P--L relation
for Type~{\small I} Cepheids breaks down for the longer period ($>
100$\,d) Cepheids -- it is not known whether the RV~Tauri stars also show
this behaviour. Complications in this investigation will be the increasing
complexity of the light curve behaviour of these longer period stars and the
possibility that larger amounts of circumstellar reddening may be present,
affecting our results in the visual bandpasses. There are indications that
some of the longer period RV~Tauri stars do show near-infrared excesses
(Welch 1987, Laney 1991).

The post-AGB phase of evolution is believed to be extremely fast and
long-term two-colour photometry on a common photometric system, such as what
has been obtained to date in the MACHO project, will be invaluable in
looking for period and colour changes due to evolutionary effects.  In
addition, this database will allow us to investigate the relatively recent
suggestion that the RV~Tauri stars show evidence of chaotic behaviour
(Buchler \& Kovacs 1987, Kovacs \& Buchler 1988, Buchler et al.\ 1995) and
the Type~{\small II} RV~Tauri sequence in the LMC will be an extremely
useful tool in exploring the predicted trend in chaotic behaviour with
$T_{\rm eff}$ (or luminosity).
   
More fields from the MACHO project microlensing survey should be
examined to increase the current sample of Type~{\small II} Cepheids and
RV~Tauri stars in the LMC. In addition a search for Type~{\small II}
Cepheids and RV~Tauri star candidates in the SMC would be useful in
investigating the effects of the lower metallicity on the group
characteristics and properties.

\acknowledgments

We would like to thank Drs Michael Albrow and Luis Balona for computer
assistance in the data analysis.
KRP was a Postdoctoral Research Fellow at the South African Astronomical
Observatory (SAAO) during this work.
The SAAO is a National Research Facility operated by the Foundation for
Research Development.  
We are very grateful for the skilled support given our project 
by the technical staff at the Mt. Stromlo Observatory.  
Work performed at LLNL is supported by the DOE under contract 
W7405-ENG-48.
Work performed by the Center for Particle Astrophysics on the UC campuses
is supported in part by the Office of Science and Technology Centers of
NSF under cooperative agreement AST-8809616.
Work performed at MSSSO is supported by the Bilateral Science 
and Technology Program of the Australian Department of Industry, 
Technology
and Regional Development. KG acknowledges a DOE OJI grant, and CWS
and KG thank the Sloan Foundation for their support.

\clearpage

\begin{deluxetable}{lcccccccccl}
\scriptsize
\tablecaption{Period and light curve analysis of the LMC Type~{\sc ii} 
Cepheid and RV~Tauri stars. \label{tabdata1} }
\tablehead{
\colhead{star id} & 
\multicolumn{2}{c}{$\alpha~~~(2000.0)~~~\delta$} & 
\colhead{Period} &
\colhead{$\log P$} & 
\colhead{$V_{\rm M}$} & 
\colhead{$\Delta V_{\rm M}$} &
\colhead{$R_{\rm M}$} & 
\colhead{$\Delta R_{\rm M}$} & 
\colhead{type\tablenotemark{a}} & 
\colhead{comments} \\
\colhead{} &
\colhead{} & 
\colhead{} & 
\colhead{(days)} & 
\colhead{} & 
\colhead{(mag)} & 
\colhead{(mag)} & 
\colhead{(mag)} & 
\colhead{(mag)} & 
\colhead{}
}  
\startdata
12.10318.52 & 05 43 47.2 & -70 42 39 &~8.240 &0.916 & -7.13& 0.59 &-7.42  &0.46 &s&\nl
81.8756.207 & 05 34 37.8 & -70 01 07 &~9.309 &0.969 & -6.94& 0.54 &-7.28  &0.45 &s&\nl    
1.3812.61   & 05 03 59.3 & -68 53 24 &~9.387 &0.973 & -7.64& 0.27 &-7.80  &0.22 &s&\nl
10.4040.38  & 05 05 11.8 & -69 45 55 &~9.622 &0.983 & -7.65& 0.43 &-7.79  &0.32 &s&\nl
80.6469.135 & 05 20 20.7 & -69 12 21 &10.509 &1.022 & -7.16& 0.36 &-7.51  &0.27 &s&\nl
80.6590.137 & 05 21 18.8 & -69 11 47 &11.442 &1.058 & -7.49& 0.92 &-7.68  &0.68 &c&\nl
3.7332.39   & 05 25 14.9 & -68 09 12 &12.704 &1.104 & -7.17& 1.02 &-7.52  &0.80 &f&\nl
6.6696.72   & 05 21 35.2 & -70 13 26 &12.902 &1.111 & -7.16& 0.63 &-7.55  &0.51 &s&\nl
10.4162.33  & 05 06 28.7 & -69 43 58 &13.246 &1.122 & -7.19& 0.42 &-7.53  &0.37 &s&\nl
80.6475.2289& 05 20 10.4 & -68 48 40 &13.925 &1.144 & -7.26& 1.30 &-7.62  &0.95 &f&\nl
81.9006.64  & 05 36 02.8 & -69 27 16 &14.337 &1.156 & -7.02& 0.95 &-7.48  &0.77 &f&\nl
47.2611.589 & 04 56 15.9 & -68 16 16 &14.469 &1.160 & -7.10& 1.14 &-7.50  &0.92 &f&\nl
19.4425.231 & 05 07 38.9 & -68 20 06 &14.752 &1.169 & -7.63& 0.87 &-7.90  &0.61 &c&\nl
2.5877.58   & 05 16 35.6 & -68 21 02 &14.855 &1.172 & -6.55& 1.03 &-7.15  &0.83 &f&\nl
1.3808.112  & 05 04 07.8 & -69 07 31 &14.906 &1.173 & -6.81& 1.04 &-7.30  &0.84 &f&\nl
14.8983.1894& 05 36 01.7 & -71 00 12 &15.391 &1.187 & -7.25& 1.08 &-7.63  &0.91 &f&\nl
2.5025.39   & 05 11 21.5 & -68 40 12 &16.602 &1.220 & -7.33& 1.10 &-7.66  &0.90 &f&\nl
9.5117.58   & 05 11 49.3 & -70 34 11 &16.747 &1.224 & -7.57& 1.12 &-7.93  &0.92 &f&\nl
10.3680.18  & 05 02 53.3 & -69 36 58 &17.127 &1.234 & -7.54& 1.04 &-7.92  &0.88 &f&\nl
78.6338.24  & 05 19 26.9 & -69 51 52 &17.560 &1.245 & -8.32& 0.57 &-8.48  &0.52 &c& HV~5756 \nl                                                                      
2.5026.30   & 05 11 33.5 & -68 35 54 &21.486 &1.332 & -7.46& 1.25 &-7.88  &0.97 &f/r& HV~13025\nl
78.6698.38  & 05 21 49.3 & -70 04 35 &24.848 &1.395 & -8.28& 0.95 &-8.64  &0.71 &c/r&\nl
77.7069.213 & 05 23 43.5 & -69 32 07 &24.935 &1.397 & -7.24& 1.27 &-7.67  &1.05 &r&\nl
82.8041.17  & 05 29 38.8 & -69 15 12 &26.594 &1.425 & -8.01& 1.01 &-8.38  &0.86 &f/r&\nl
81.9362.25  & 05 37 45.0 & -69 54 16 &30.408 &1.483 & -7.77& 1.22 &-8.26  &1.04 &r&\nl
14.9582.9   & 05 39 33.1 & -71 21 55 &31.127 &1.493 & -8.64& 0.94 &-8.91  &0.75 &r& HV~12631 \nl
19.3694.19  & 05 03 05.0 & -68 40 25 &31.716 &1.501 & -8.88& 0.93 &-9.10  &0.72 &r& HV~2281 \nl
78.5856.2363& 05 16 47.5 & -69 44 15 &41.118 &1.614 & -8.94& 1.00 &-9.19  &0.76 &r& HV~2423 \nl
81.8520.15  & 05 32 54.5 & -69 35 13 &42.079 &1.624 & -9.29& 0.39 &-9.42  &0.29 &s?&\nl
82.8405.15  & 05 31 50.9 & -69 11 46 &46.542 &1.668 & -9.78& 0.70 &-9.86  &0.54 &r&\nl
81.9728.14  & 05 40 00.5 & -69 42 14 &47.019 &1.672 & -9.63& 0.40 &-9.95  &0.26 &r?&\nl
79.5501.13  & 05 14 18.1 & -69 12 35 &48.539 &1.686 & -9.52& 0.79 &-9.71  &0.65 &r& HV~915\nl
47.2496.8   & 04 55 43.2 & -67 51 10 &56.224 &1.750 & -9.03& 0.90 &-9.48  &0.68     &r&\nl 
\enddata
\tablenotetext{a}{s=sinusoidal, c=crested, f=flat-topped, r=RV~Tauri like, 
?=uncertain classification}
\end{deluxetable} 

\clearpage

{\footnotesize

\begin{deluxetable}{lccccccccc}
\scriptsize
\tablecaption{Period and luminosity data for the LMC Type~{\sc ii} Cepheid and 
RV~Tauri stars. \label{tabdata2}}
\tablehead{
\colhead{star id} & 
\colhead{Period} & 
\colhead{$\log P$} & 
\colhead{$V$} &
\colhead{ $E(B-V)$} & 
\colhead{$V_{\circ}$} &
\colhead{$R_{\circ}$} & 
\colhead{$(V-R)_{\circ}$} & 
\colhead{$W_{V}$} &
\colhead{$M_{V}$} \\
\colhead{} & 
\colhead{days} &
\colhead{} & 
\colhead{(mag)} & 
\colhead{(mag)} & 
\colhead{(mag)} & 
\colhead{(mag)} & 
\colhead{(mag)} & \\ 
}
\startdata
12.10318.52 &~8.240 &0.916 & 17.07 & 0.15 & 16.61  & 16.32 &  0.291& 15.01 & -1.89 \\      
81.8756.207 &~9.309 &0.969 & 17.17 & 0.12 & 16.80  & 16.45 &  0.348& 14.89 & -1.70 \\    
1.3812.61   &~9.387 &0.973 & 16.31 & 0.13 & 15.90  & 15.72 &  0.182& 14.90 & -2.60 \\
10.4040.38  &~9.622 &0.983 & 16.51 & 0.14 & 16.08  & 15.88 &  0.198& 14.99 & -2.42 \\
80.6469.135 &10.509 &1.022 & 17.05 & 0.09 & 16.77  & 16.37 &  0.394& 14.60 & -1.73 \\
80.6590.137 &11.442 &1.058 & 16.72 & 0.08 & 16.47  & 16.18 &  0.297& 14.84 & -2.03 \\
3.7332.39   &12.704 &1.104 & 17.00 & 0.07 & 16.79  & 16.40 &  0.387& 14.66 & -1.71 \\
6.6696.72   &12.902 &1.111 & 16.86 & 0.12 & 16.49  & 16.15 &  0.341& 14.62 & -2.01 \\
10.4162.33  &13.246 &1.122 & 16.95 & 0.14 & 16.52  & 16.19 &  0.329& 14.71 & -1.98 \\
80.6475.2289&13.925 &1.144 & 16.75 & 0.08 & 16.50  & 16.13 &  0.370& 14.47 & -2.00 \\
81.9006.64  &14.337 &1.156 & 16.88 & 0.09 & 16.61  & 16.25 &  0.358& 14.64 & -1.89 \\
47.2611.589 &14.469 &1.160 & 16.89 & 0.13 & 16.49  & 16.13 &  0.355& 14.54 & -2.01 \\
19.4425.231 &14.752 &1.169 & 16.16 & 0.13 & 15.75  & 15.53 &  0.222& 14.54 & -2.75 \\
2.5877.58   &14.855 &1.172 & 17.31 & 0.11 & 16.97  & 16.50 &  0.464& 14.42 & -1.53 \\
1.3808.112  &14.906 &1.173 & 17.26 & 0.14 & 16.82  & 16.40 &  0.423& 14.50 & -1.68 \\
14.8983.1894&15.391 &1.187 & 16.90 & 0.14 & 16.46  & 16.10 &  0.364& 14.46 & -2.04 \\
2.5025.39   &16.602 &1.220 & 16.58 & 0.13 & 16.18  & 15.87 &  0.308& 14.49 & -2.32 \\
9.5117.58   &16.747 &1.224 & 16.49 & 0.14 & 16.05  & 15.73 &  0.318& 14.30 & -2.45 \\
10.3680.18  &17.127 &1.234 & 16.58 & 0.14 & 16.15  & 15.82 &  0.329& 14.34 & -2.35 \\
78.6338.24  &17.560 &1.245 & 15.68 & 0.08 & 15.43  & 15.19 &  0.239& 14.12 & -3.07 \\
2.5026.30   &21.486 &1.332 & 16.44 & 0.13 & 16.03  & 15.67 &  0.361& 14.05 & -2.47 \\
78.6698.38  &24.848 &1.395 & 15.69 & 0.12 & 15.25  & 14.93 &  0.323& 13.48 & -3.25 \\
77.7069.213 &24.935 &1.397 & 16.57 & 0.08 & 16.32  & 15.90 &  0.427& 13.98 & -2.18 \\
82.8041.17  &26.594 &1.425 & 15.99 & 0.08 & 15.74  & 15.34 &  0.399& 13.55 & -2.76 \\
81.9362.25  &30.408 &1.483 & 16.09 & 0.12 & 15.72  & 15.34 &  0.377& 13.65 & -2.78 \\
14.9582.9   &31.127 &1.493 & 15.48 & 0.12 & 15.10  & 14.84 &  0.259& 13.68 & -3.40 \\
19.3694.19  &31.716 &1.501 & 15.29 & 0.13 & 14.89  & 14.64 &  0.248& 13.53 & -3.61 \\
78.5856.2363&41.118 &1.614 & 15.10 & 0.12 & 14.72  & 14.44 &  0.288& 13.14 & -3.78 \\
81.8520.15  &42.079 &1.624 & 14.64 & 0.10 & 14.33  & 14.20 &  0.124& 13.65 & -4.17 \\
82.8405.15  &46.542 &1.668 & 14.26 & 0.09 & 13.98  & 13.77 &  0.208& 12.83 & -4.52 \\
81.9728.14  &47.019 &1.672 & 14.31 & 0.10 & 14.00  & 13.74 &  0.267& 12.54 & -4.50 \\
79.5501.13  &48.539 &1.686 & 14.48 & 0.12 & 14.11  & 13.89 &  0.215& 12.93 & -4.39 \\ 
47.2496.8   &56.224 &1.750 & 14.97 & 0.11 & 14.63  & 14.23 &  0.394& 12.46 & -3.87 \\
\enddata 
\end{deluxetable} 

\clearpage

\begin{deluxetable}{lccl}
\tablecaption{Period--luminosity relationships for Type~{\sc ii} Cepheids. 
\label{tabpl}}
\tablehead{
\colhead{relation} & 
\colhead{validity} & 
\colhead{note} & 
\colhead{reference} \\
\colhead{$M_{V}=$} & 
\colhead{} &
\colhead{} & 
\colhead{} \\ 
}
\startdata
$-0.13 - 1.90 \log P$ & $ 0.0 < \log P < 1.3$  & & Dickens \& Carey (1967)\nl
$-5.3~ + 0.042 P$     & $1.47 < \log P < 1.72$ & & DuPuy (1973) \nl
$-0.08 - 1.59 \log P$ & $ 0.0 < \log P < 1.3$   & & Demers \& Harris (1974)\nl
$-0.08 - 1.59 \log P$ & $ 0.0 < \log P < 1.1$   & & Harris (1981) \nl
$+2.13 - 3.60 \log P$ & $ 1.1 < \log P < 1.6$   & & Harris (1981) \nl
$+0.20 - 1.93 \log P$ & $-0.2 < \log P < 1.53$  & \tablenotemark{b}
& Nemec et al.\ (1994) \nl
$-0.25 - 1.93 \log P$ & $-0.2 < \log P < 1.53$  & \tablenotemark{c}  
& Nemec et al.\ (1994) \nl 
$+1.34 - 3.07 \log P$ & $0.9 < \log P < 1.75$   & & this study \nl
$-0.61 - 2.95 \log P$ + 5.49$(V-R)_{\circ}$ & $0.9 < \log P < 1.75$ 
& & this study \nl 
\enddata
\tablenotetext{b}{assuming [Fe/H] = --2 and fundamental mode of pulsation} 
\tablenotetext{c}{assuming [Fe/H] = --2 and first-harmonic mode of pulsation}
\end{deluxetable}

\clearpage

\begin{center}
{FIGURE CAPTIONS}
\end{center}

\figcaption[]{The finding charts for the thirty three MACHO Type~{\sc ii}
Cepheids and RV~Tauri stars in the LMC. The target stars are indicated with 
crosshairs. North is to the top and east is to the left. The maximum size of
the charts is 3 arcmin $\times$ 3 arcmin.\label{figfinder}}

\figcaption[]{The $V_{\rm M}$ light curves for the LMC Type~{\sc ii} Cepheids 
and RV~Tauri stars described in this paper. The stars with periods
less then 20\,d (the Type~{\sc ii} Cepheids) are plotted on the `single' or
fundamental period, while the stars with periods greater than 20\,d
(generally RV~Tauri stars) are plotted on the `double' or formal period. 
These periods are indicated at the bottom of each panel. The variables are
plotted between phases 0.0 and 1.5 in order to reveal continuity. The large
ticks on the vertical axis correspond to 0.5
magnitudes.\label{figlightcurves}}

\figcaption[]{The $(V-R)_{\rm M}$ colour curves for the 
Type~{\sc ii} Cepheids and RV~Tauri stars in the LMC.  Periods and epochs
are as for Figure~\protect\ref{figlightcurves}. The large ticks on the
vertical axis correspond to 0.2 magnitudes.\label{figcolourcurves}}

\figcaption[]{The $V_{\rm M}$ light curve for  MACHO*05:37:45.0-69:54:16 
plotted together with the 5th order fourier fit to the photometric data.
This star has a `formal' period of 60.816\,d.\label{figffit}}

\figcaption[]{The fourier parameters $\Phi_{21}$, $R_{21}$, $\Phi_{31}$ and 
$R_{31}$ are plotted {\it vs} $\log P$. Type~{\sc ii} Cepheids are plotted 
using symbols corresponding to their light curve appearance: sinusoidal
(crosses), crested (open triangles) and flat-topped (open squares). 
RV~Tauri stars are plotted using open circles and stars of uncertain
classification (see text) are plotted using asterisks.\label{figfourier}}

\figcaption[]{The HR diagram for the Type~{\sc ii} Cepheids and RV~Tauri stars in
the LMC.  Symbols are as for Fig.~\protect\ref{figperlac}.  The RV~Tauri stars are brighter than the
Type~{\sc ii} Cepheids for a particular intrinsic colour. The bluest stars
tend to be the short period Type~{\sc ii} Cepheids and the longest period 
RV~Tauri stars.\label{fighr}}

\figcaption[]{The a) period--luminosity, b) period--amplitude and c)
period--colour diagrams for the Type~{\sc ii} Cepheids and
RV~Tauri stars in the LMC. The luminosity is the mean $V_{\rm M}$ magnitude,
the amplitude is the peak-to-peak amplitude determined from the fourier fit
to the $V_{\rm M}$ light curve and the colour is the mean $(V-R)_{\rm M}$
colour. Symbols are as for Fig.~\protect\ref{figfourier}.\label{figperlac}}

\figcaption[]{Absolute luminosities of the LMC Type~{\sc ii} Cepheids and
RV~Tauri stars (circles) from the MACHO photometric database are plotted
{\it vs} $\log P$.  (a) Period--luminosity relations for galactic Type~{\sc
ii} Cepheids are displayed. Solid lines: Nemec et al.\ (1994) fundamental
(lower line) and harmonic (upper line) sequences assuming [Fe/H] = --2;
dotted line: Harris (1985); dashed line: DuPuy (1973).  (b)
Period--luminosity relation for the LMC Type~{\sc ii} Cepheids and RV~Tauri
stars from MACHO observations. Stars that were omitted from the analysis
(see text) are indicated by triangles.\label{figperlumold}}

\figcaption[]{Period--luminosity--colour relation for the Type~{\sc ii} Cepheids and
RV~Tauri stars in the LMC. The quantity ($V_{\circ} -
5.49\overline{(V-R)_{\circ}}$) is plotted {\it vs} $\log P$.  Type~{\sc ii}
Cepheids and RV Tauri stars (circles) are plotted using the `single' or
fundamental period.  The solid line is the direct regression relation while
the dotted line is the inverse regression relation. Stars that were omitted
in deriving the P--L--C relation (see text) are indicated by 
triangles.\label{figpl}}

\clearpage

\begin{figure*}
\begin{center}
\begin{minipage}{176mm}
\epsfxsize=176mm
\leavevmode
\epsfbox{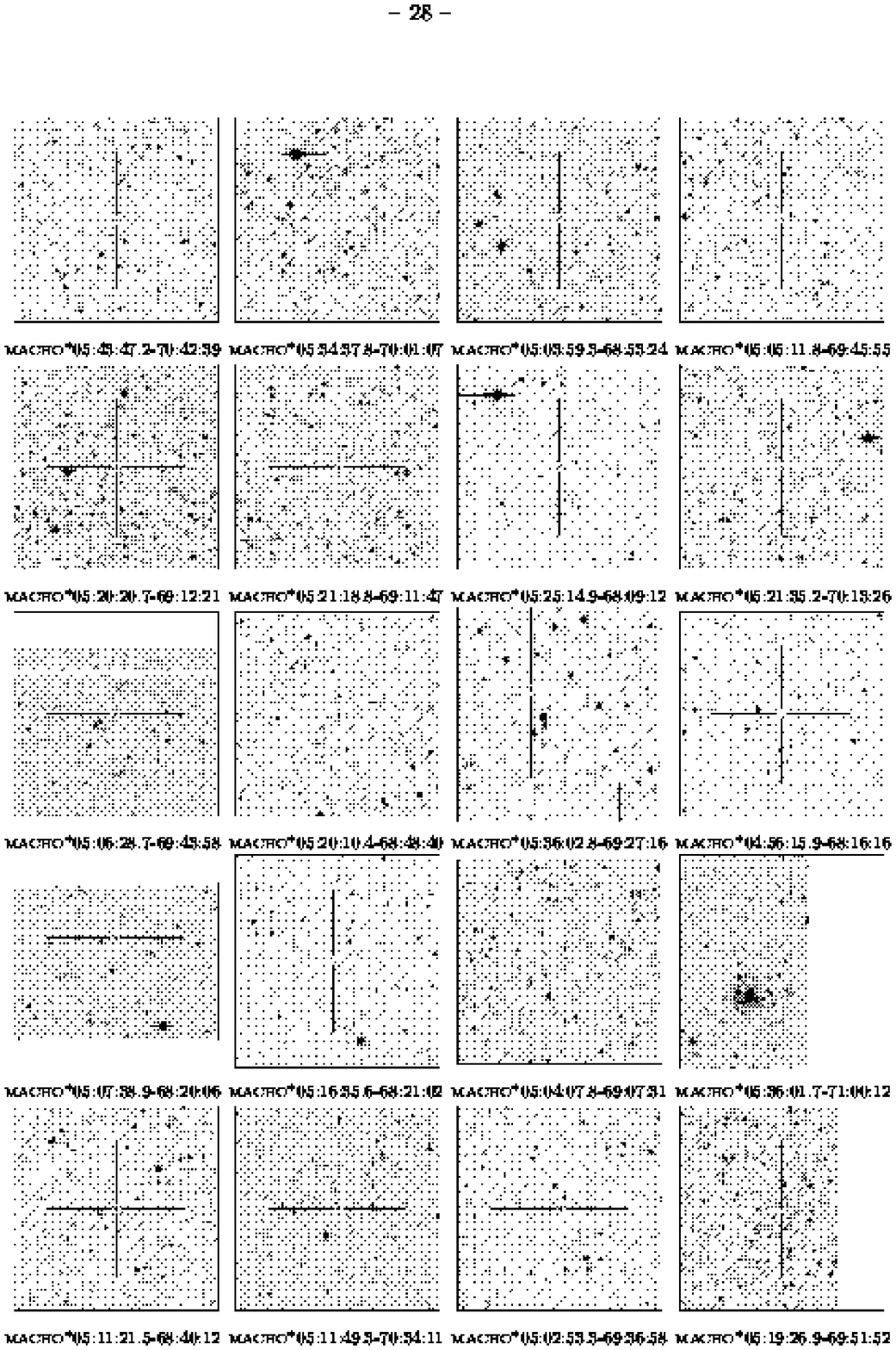}
\end{minipage}
\end{center}
\end{figure*}

\clearpage

\setcounter{figure}{0}
\begin{figure*}
\begin{center}
\begin{minipage}{176mm}
\epsfxsize=176mm
\leavevmode
\epsfbox{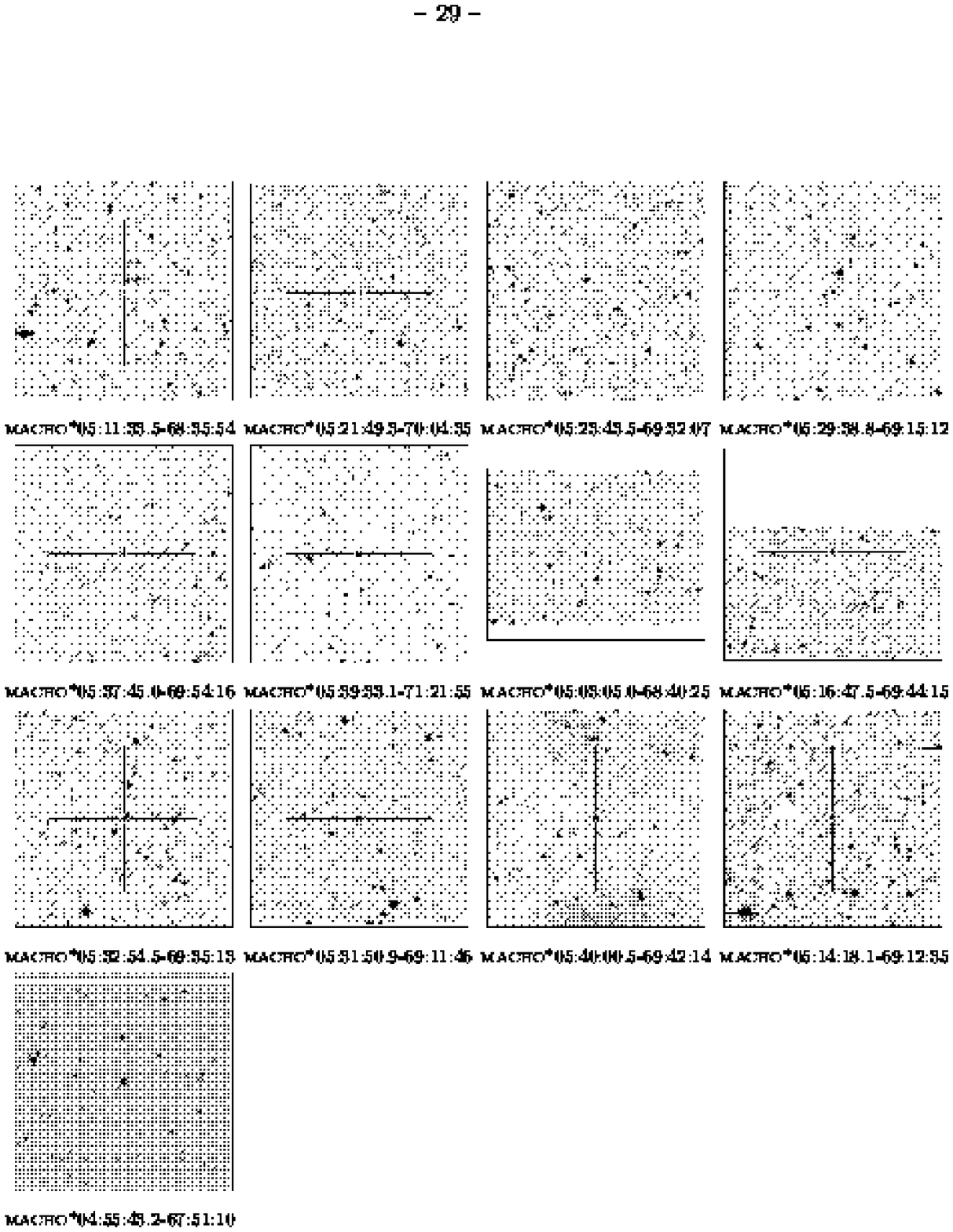}
\end{minipage}
\end{center}
\end{figure*}

\clearpage

\begin{figure*}
\begin{minipage}{176mm}
\epsfxsize=176mm
\leavevmode
\epsfbox{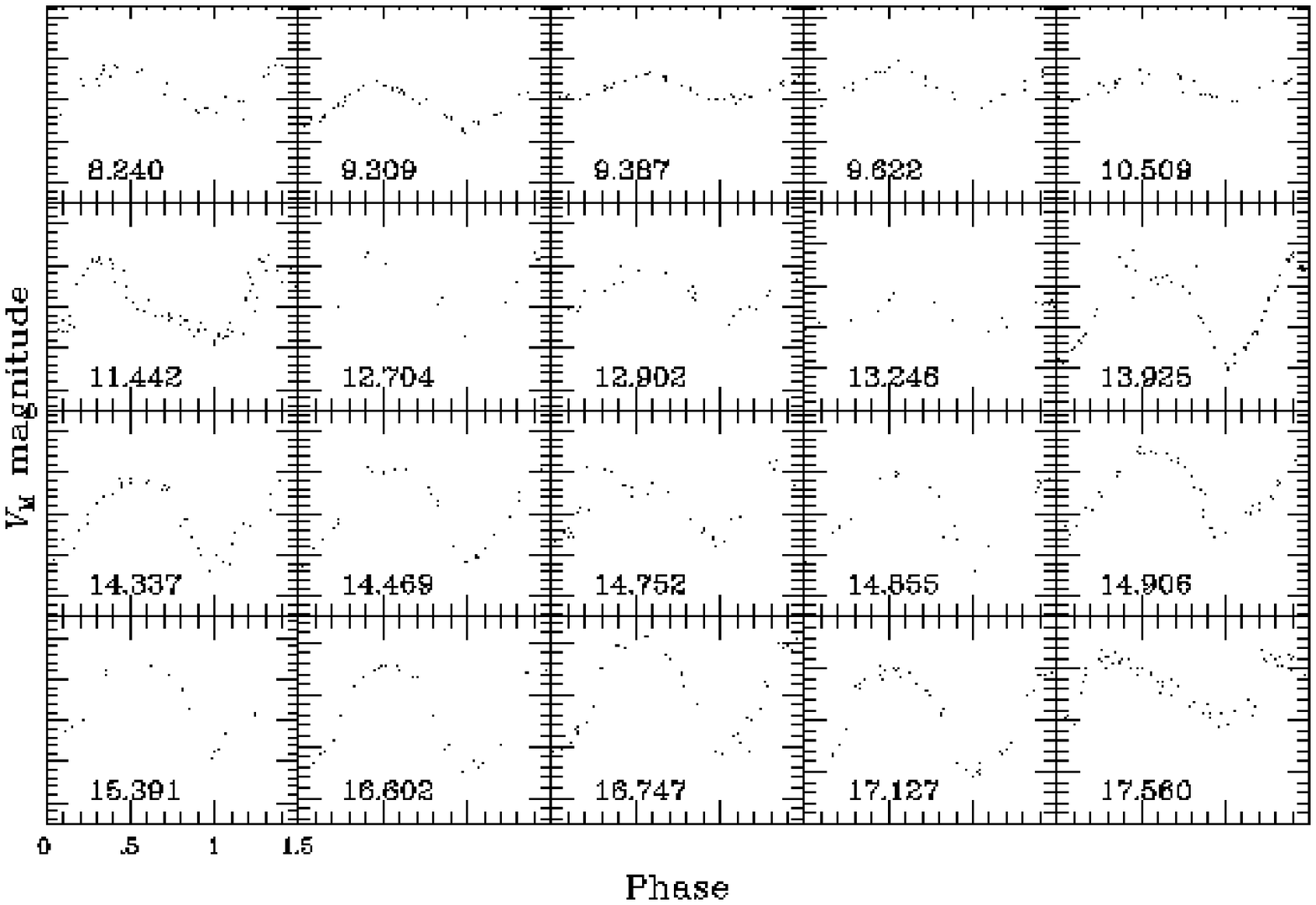}\\
\epsfxsize=176mm
\leavevmode
\epsfbox{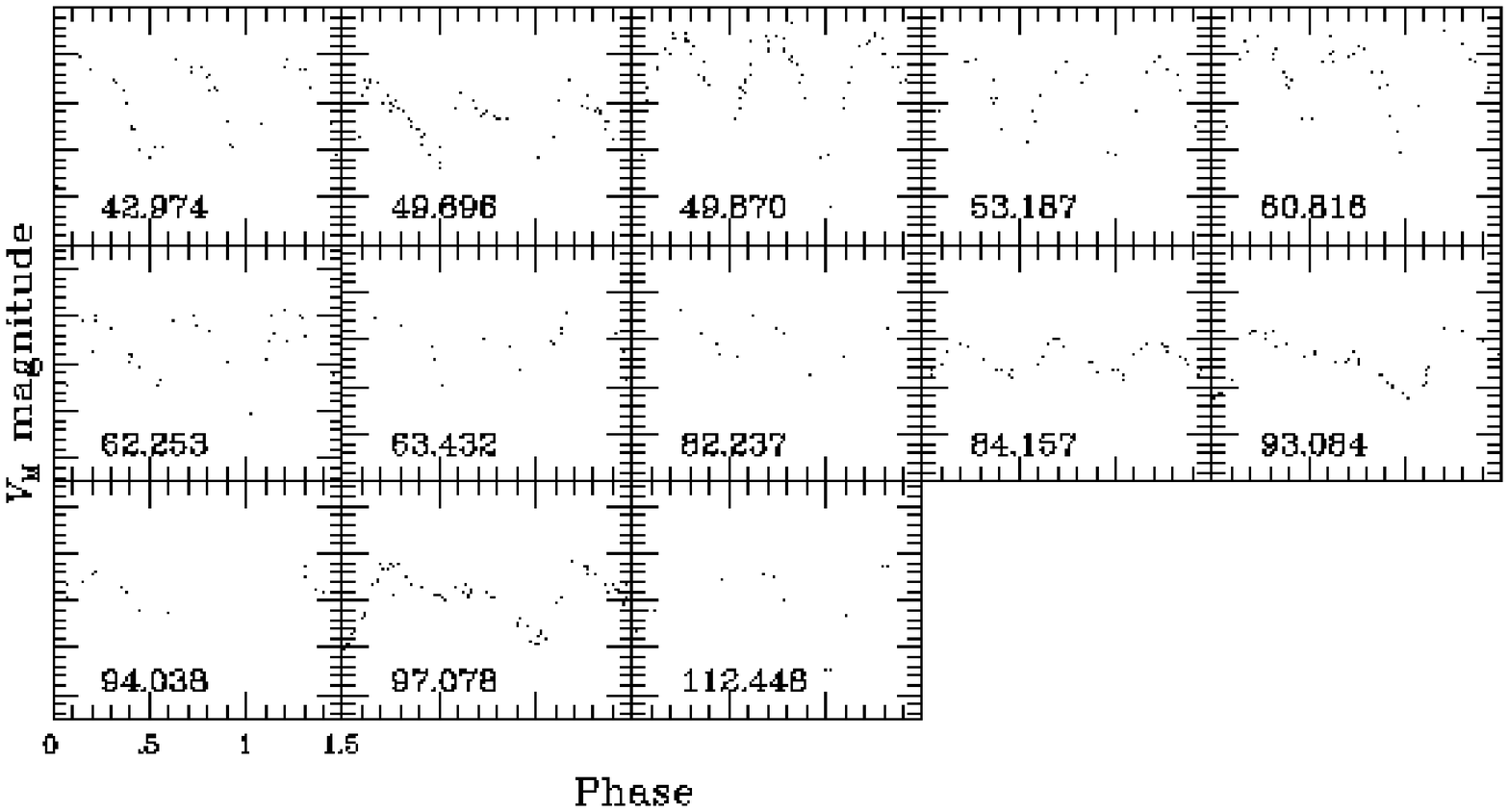}
\end{minipage}
\end{figure*}

\clearpage

\begin{figure*}
\begin{minipage}{176mm}
\epsfxsize=176mm
\leavevmode
\epsfbox{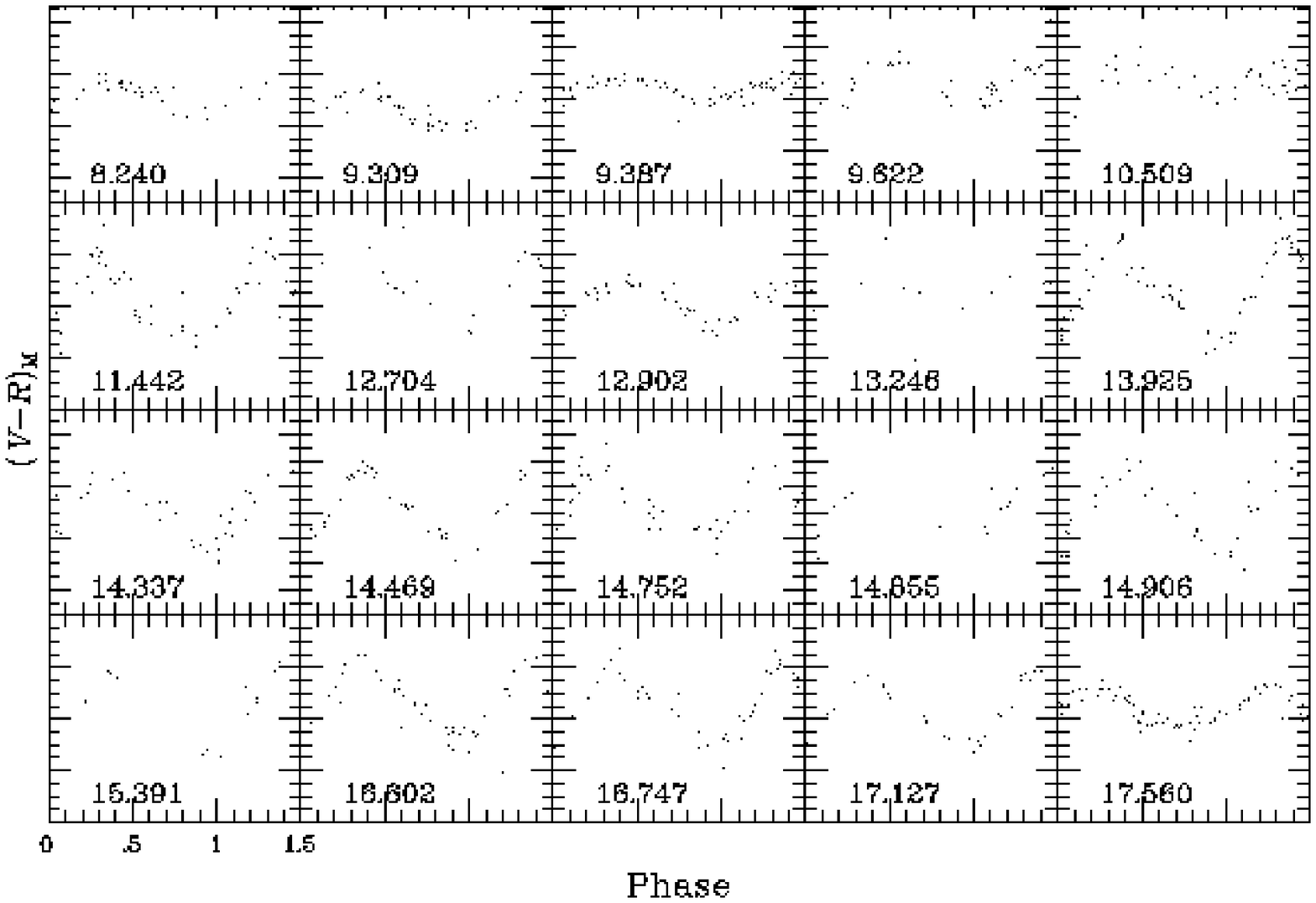}\\
\epsfxsize=176mm
\leavevmode
\epsfbox{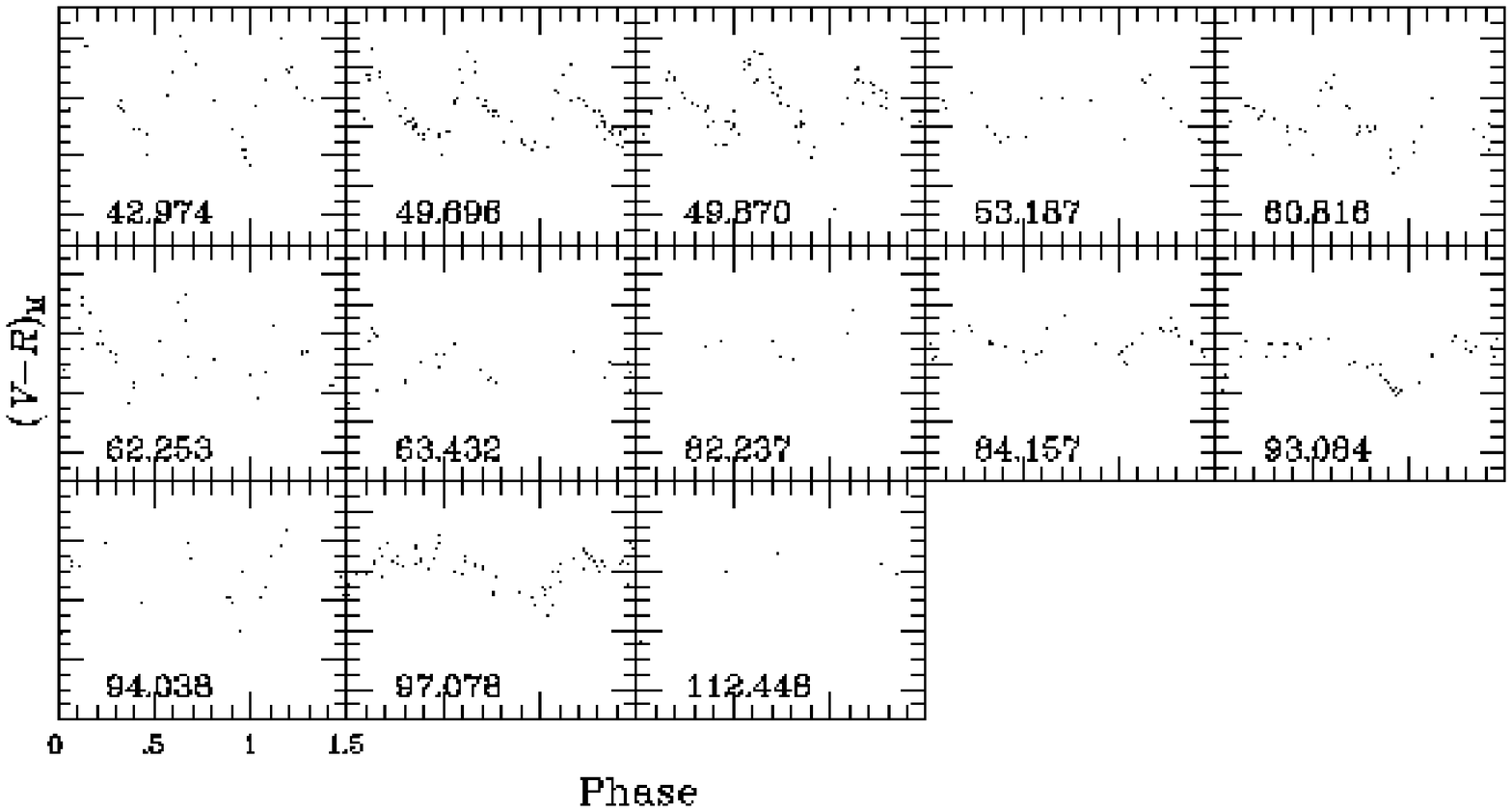}
\end{minipage}
\end{figure*}

\clearpage

\begin{figure*}
\begin{center}
\begin{minipage}{125mm}
\epsfxsize=125mm
\leavevmode
\epsfbox{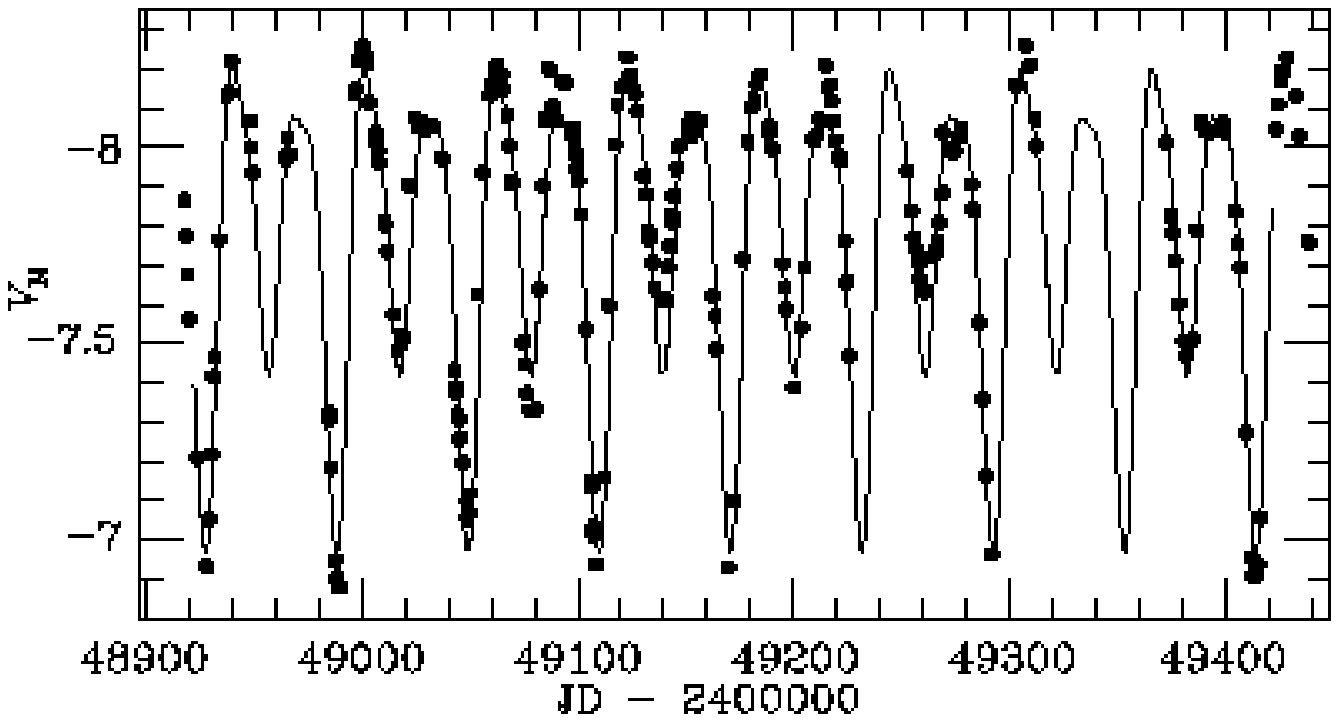}
\end{minipage}\\
\begin{minipage}{125mm}
\epsfxsize=125mm
\leavevmode
\epsfbox{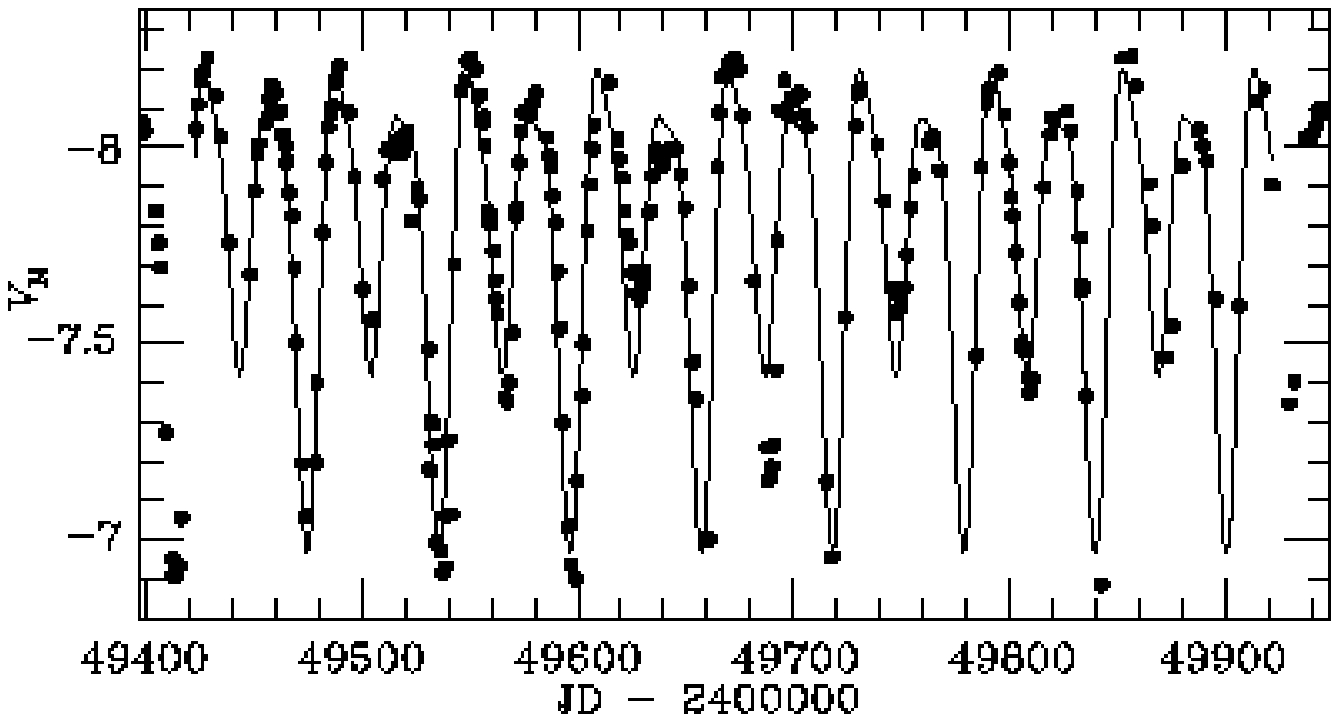}
\end{minipage}\\
\begin{minipage}{125mm}
\epsfxsize=125mm
\leavevmode
\epsfbox{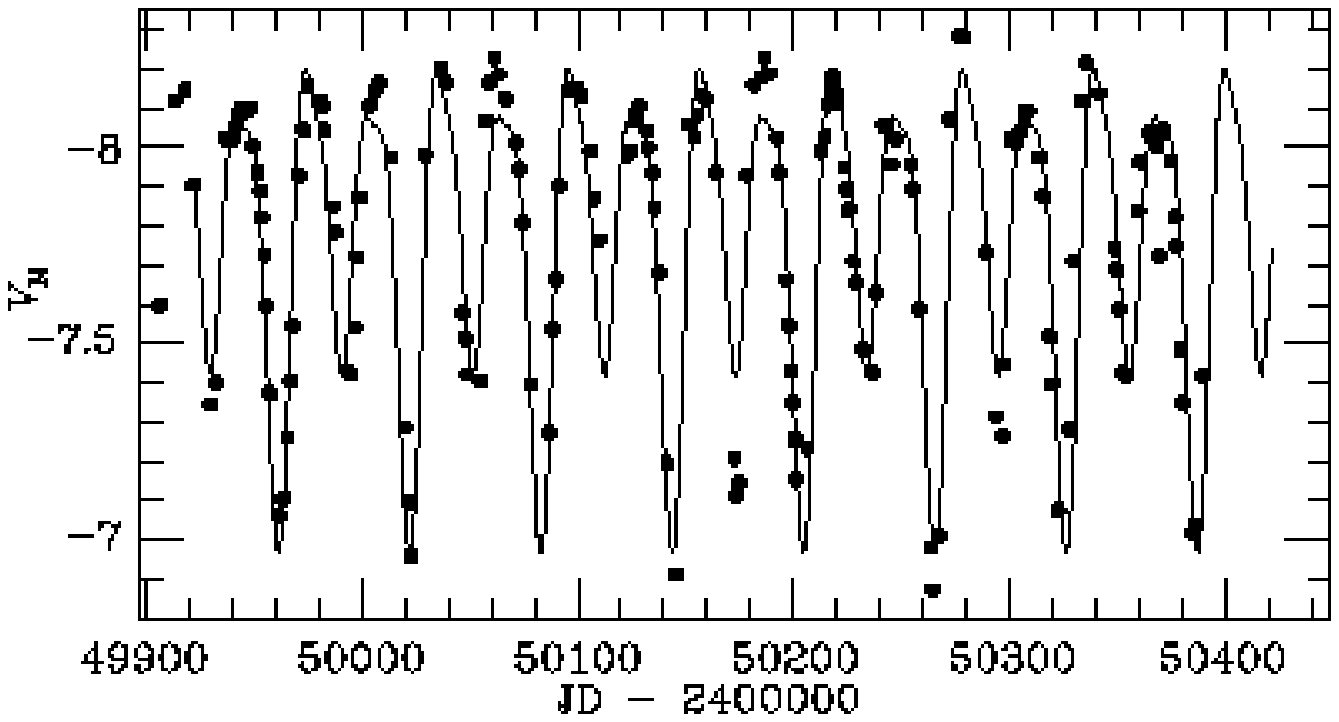}
\end{minipage}
\end{center}
\end{figure*}

\clearpage

\begin{figure*}
\begin{center}
\begin{minipage}{100mm}
\epsfxsize=100mm
\leavevmode
\epsfbox{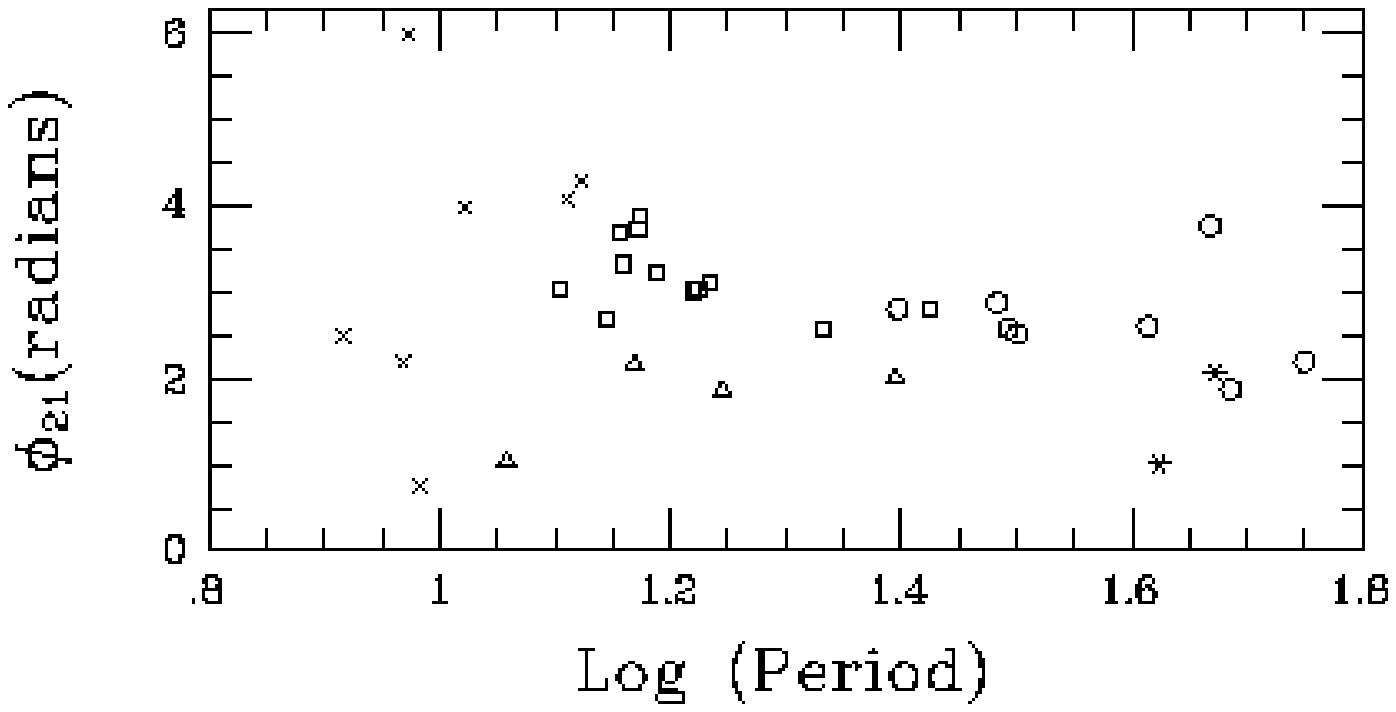}
\end{minipage}\\
\begin{minipage}{100mm}
\epsfxsize=100mm
\leavevmode
\epsfbox{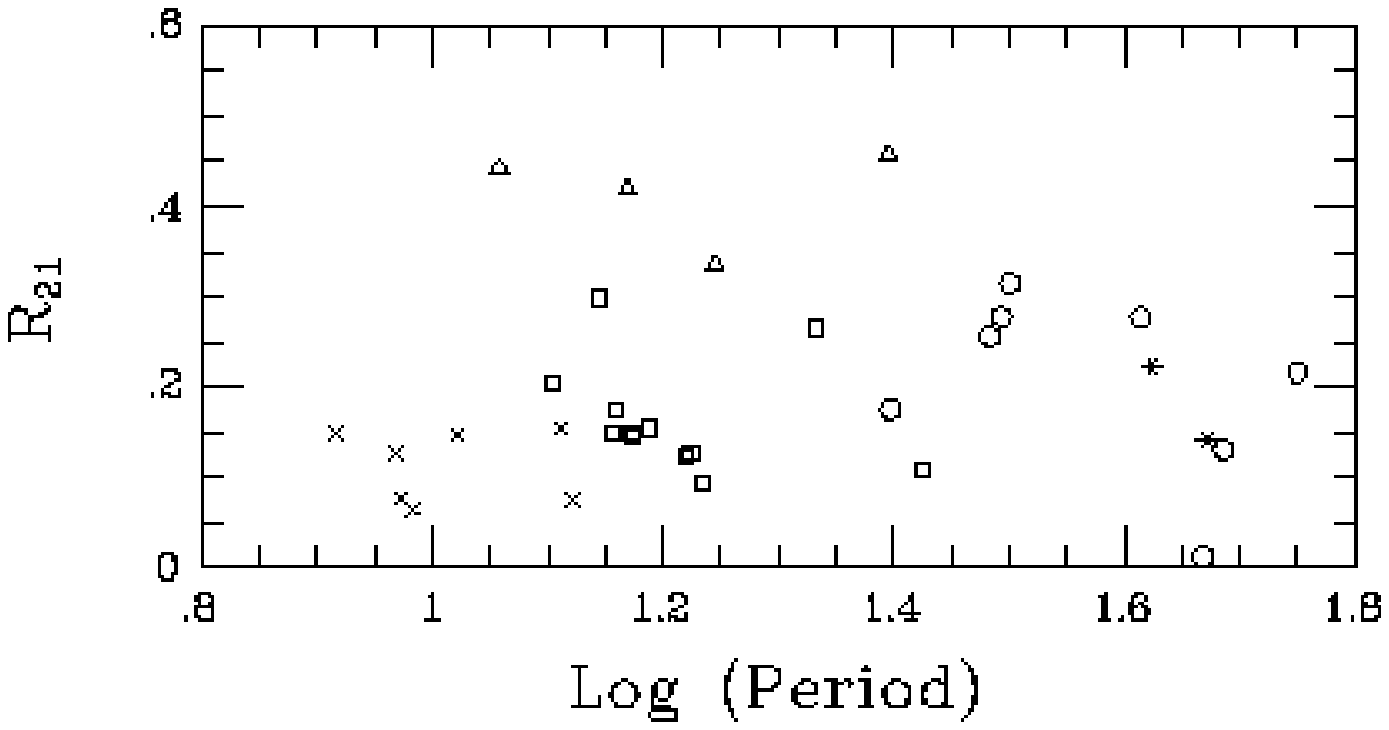}
\end{minipage}\\
\begin{minipage}{100mm}
\epsfxsize=100mm
\leavevmode
\epsfbox{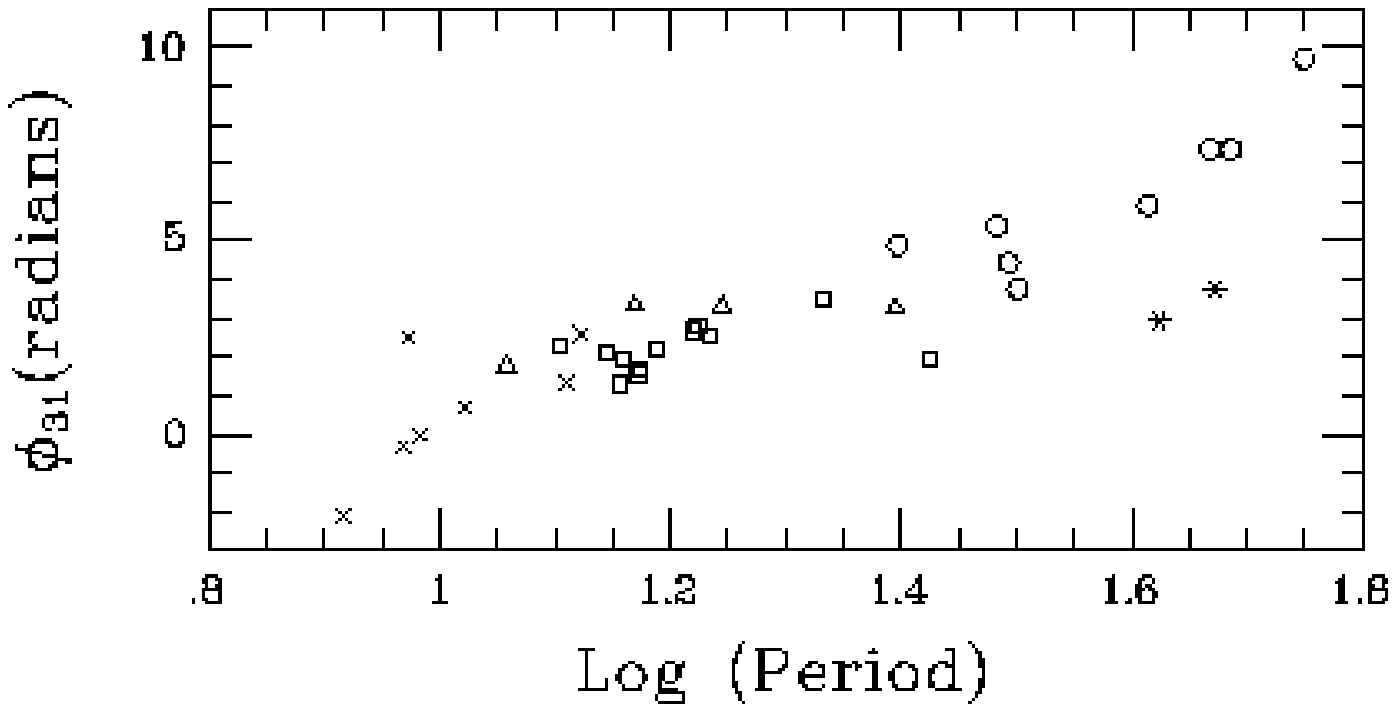}
\end{minipage}\\
\begin{minipage}{100mm}
\epsfxsize=100mm
\leavevmode
\epsfbox{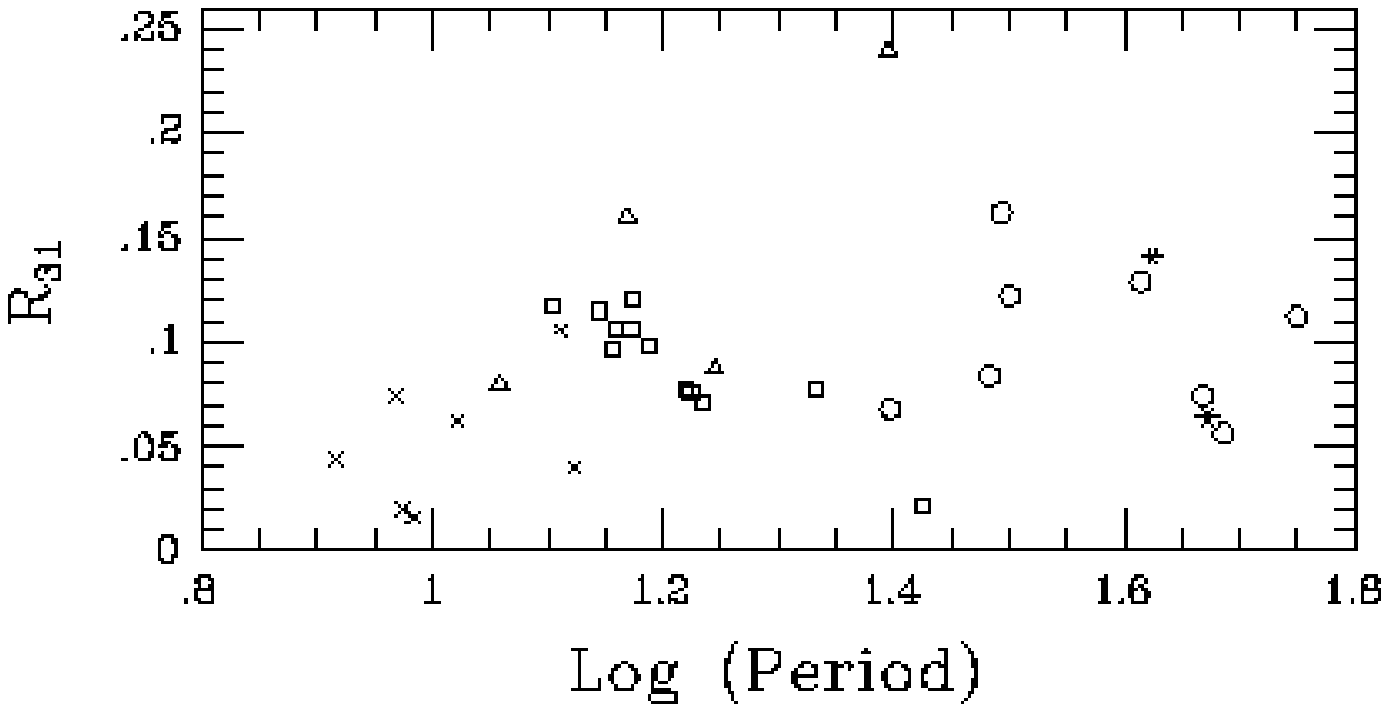}
\end{minipage}
\end{center}
\end{figure*}

\clearpage

\begin{figure*}
\begin{minipage}{150mm}
\epsfxsize=150mm
\leavevmode
\epsfbox{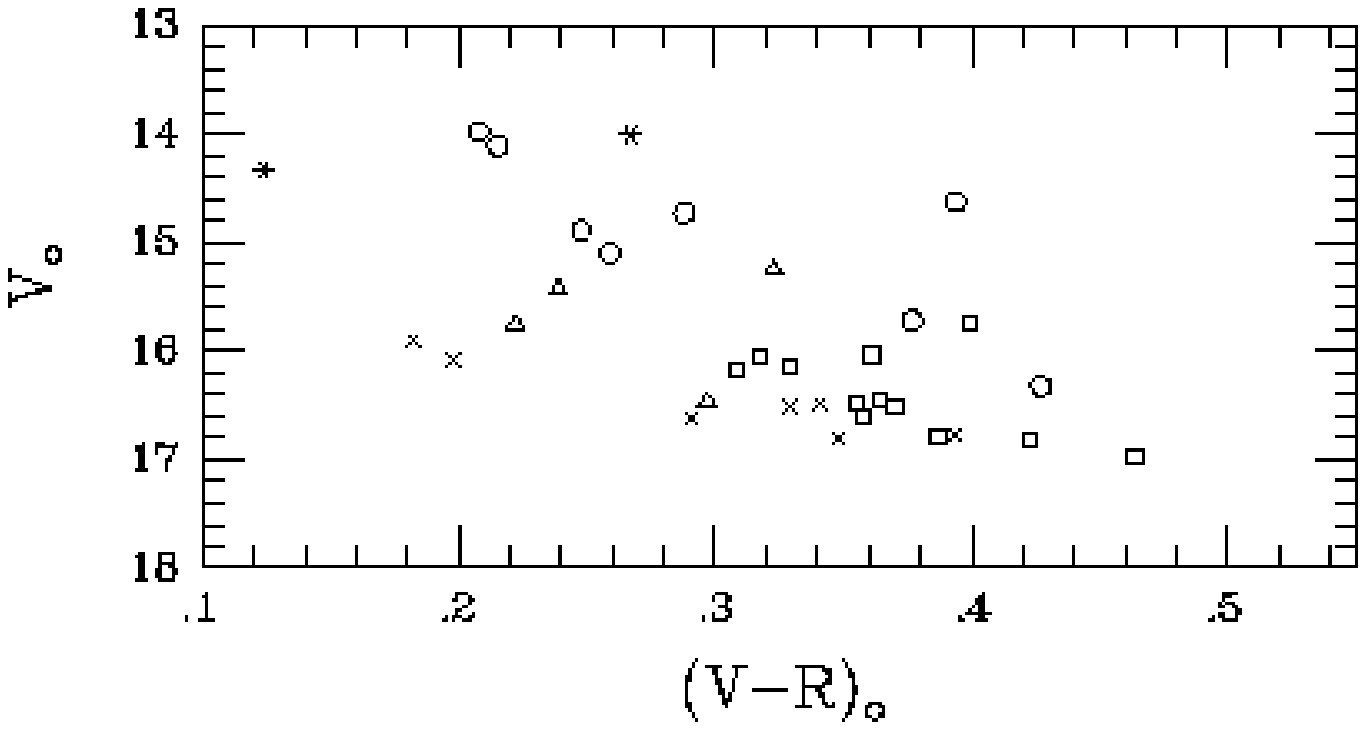}
\end{minipage}
\end{figure*}

\clearpage

\begin{figure*}
\begin{center}
\begin{minipage}{125mm}
\epsfxsize=125mm
\leavevmode
\epsfbox{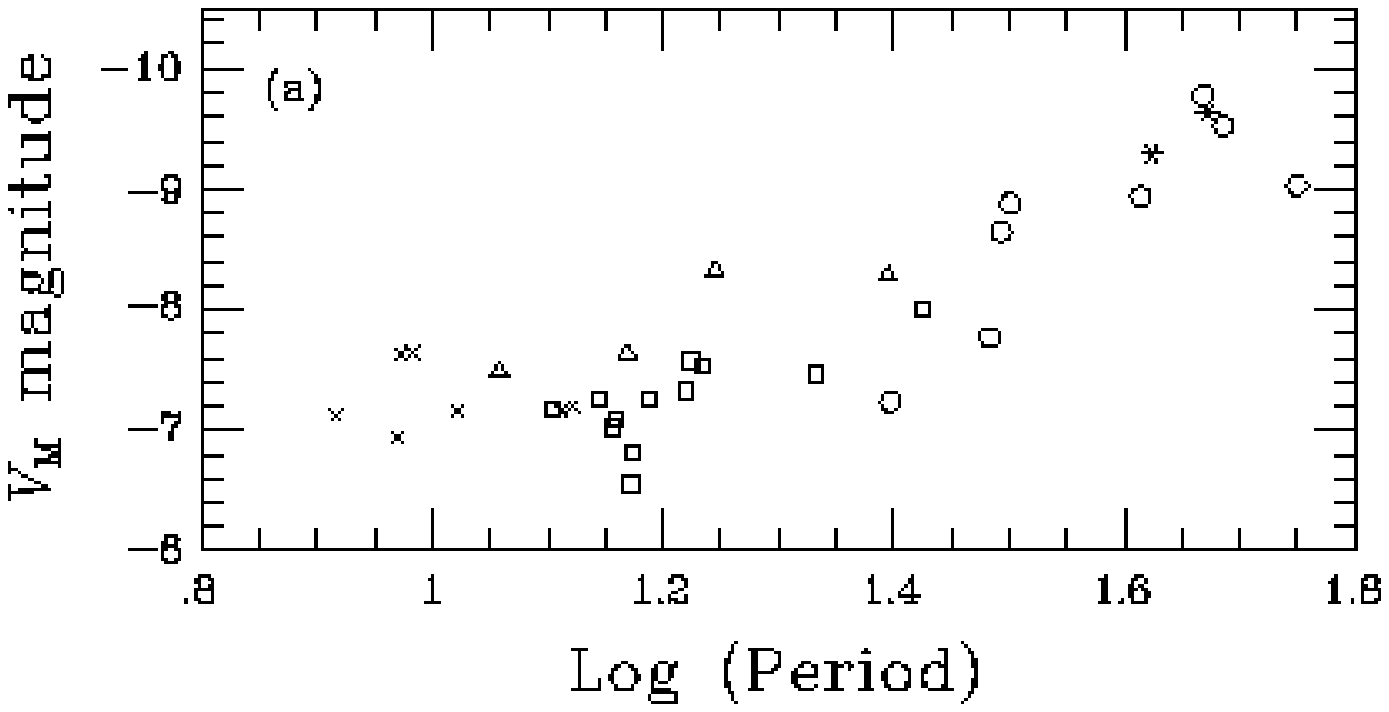}
\end{minipage}\\
\begin{minipage}{125mm}
\epsfxsize=125mm
\leavevmode
\epsfbox{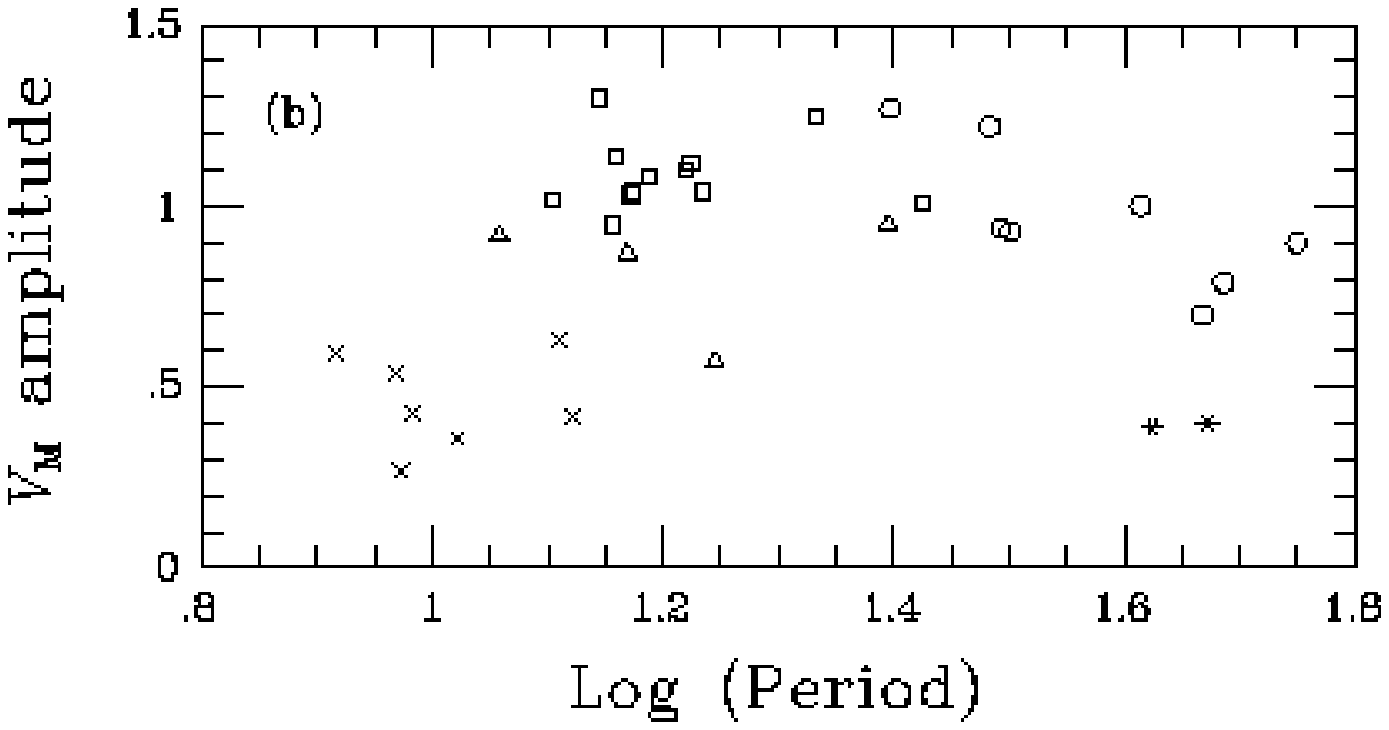}
\end{minipage}\\
\begin{minipage}{125mm}
\epsfxsize=125mm
\leavevmode
\epsfbox{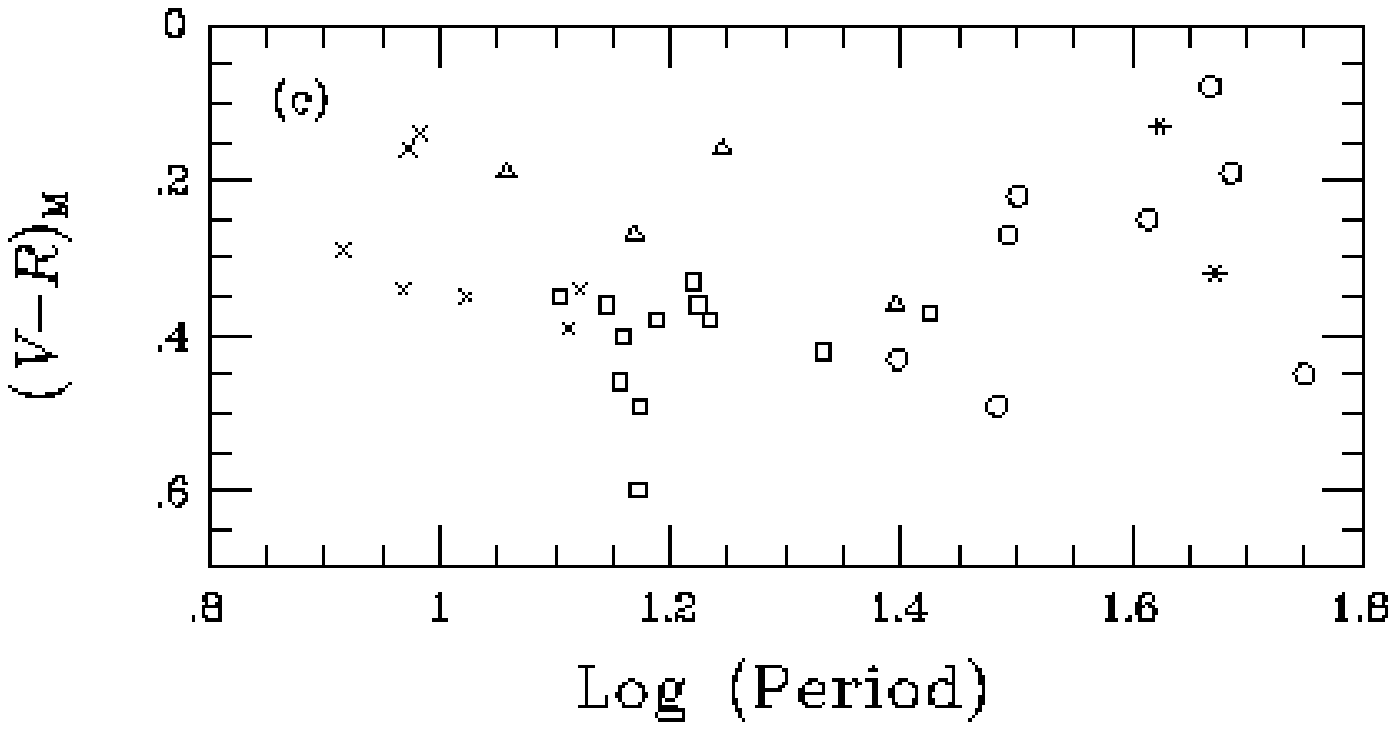}
\end{minipage}
\end{center}
\end{figure*}

\clearpage

\begin{figure*}
\begin{minipage}{150mm}
\epsfxsize=150mm
\leavevmode
\epsfbox{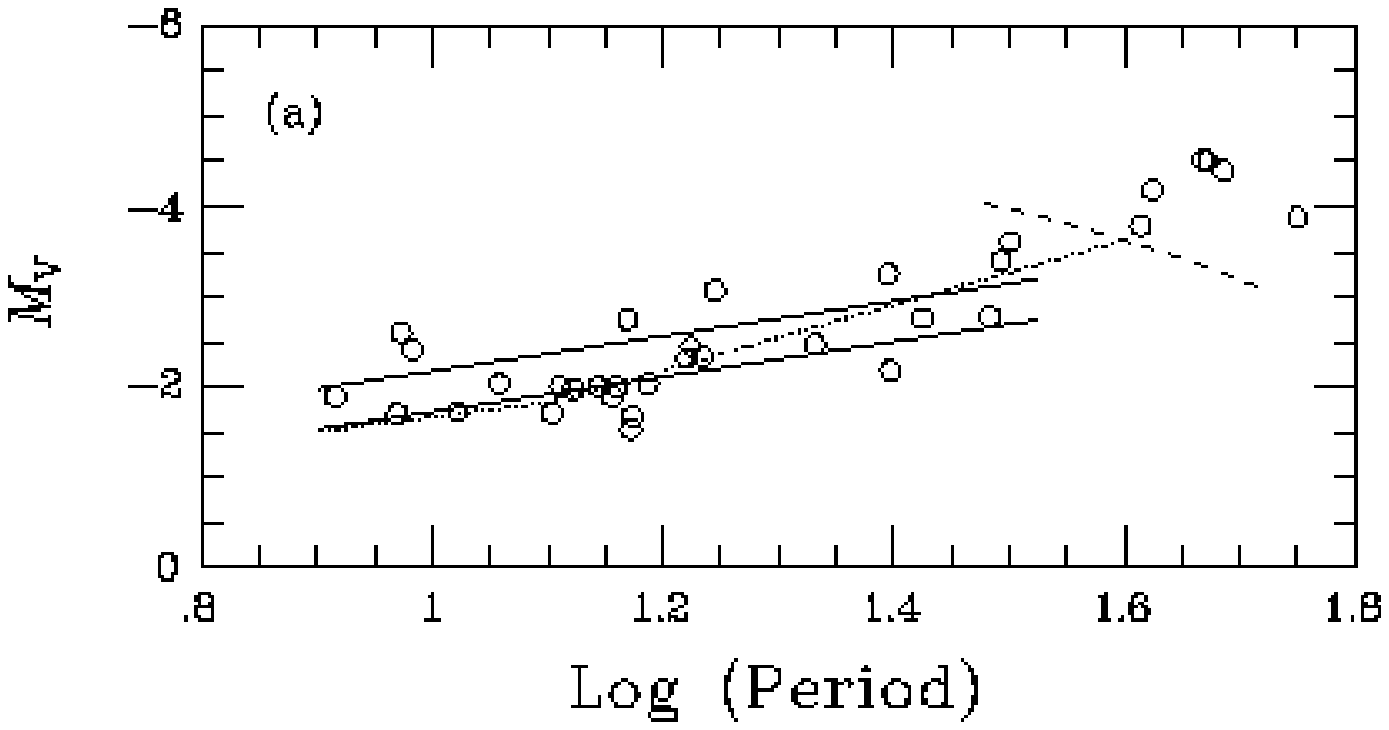}
\end{minipage}\\
\begin{minipage}{150mm}
\epsfxsize=150mm
\leavevmode
\epsfbox{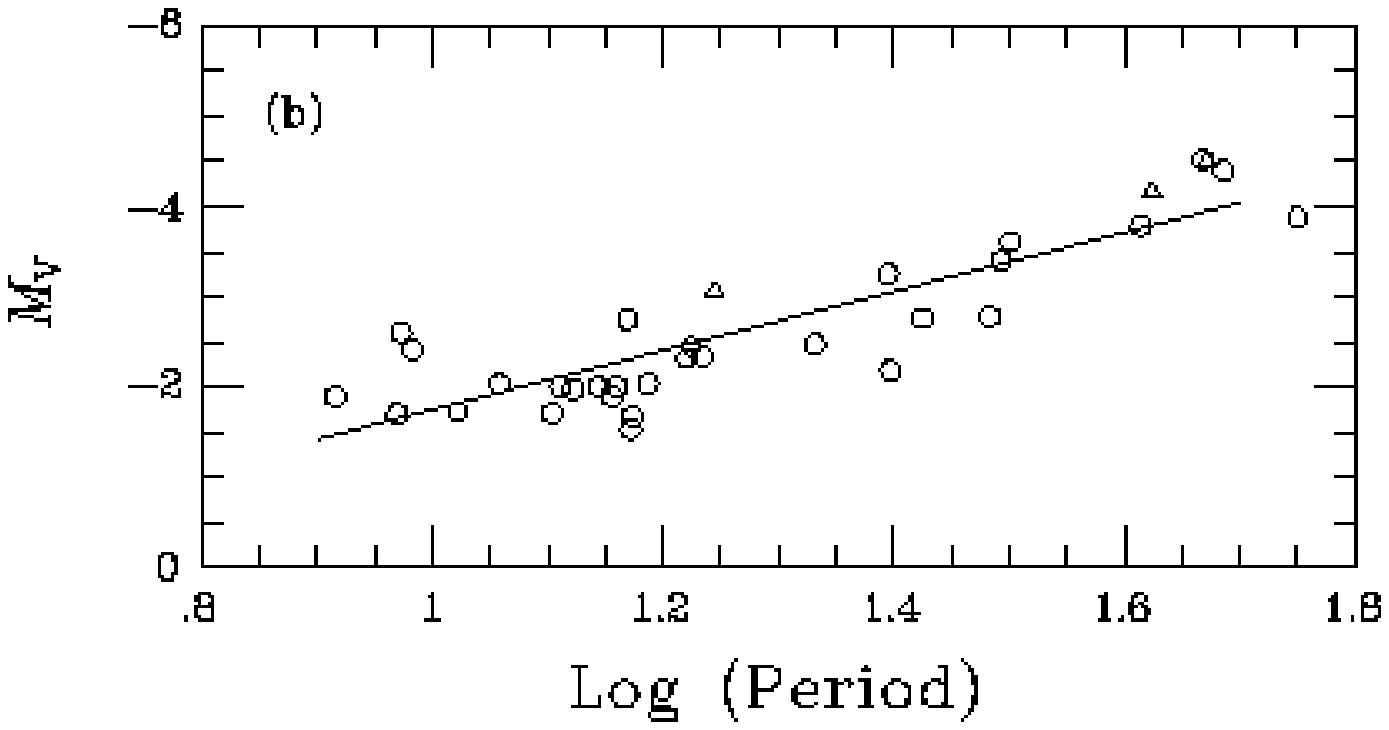}
\end{minipage}
\end{figure*}

\clearpage

\begin{figure*}
\begin{minipage}{150mm}
\epsfxsize=150mm
\leavevmode
\epsfbox{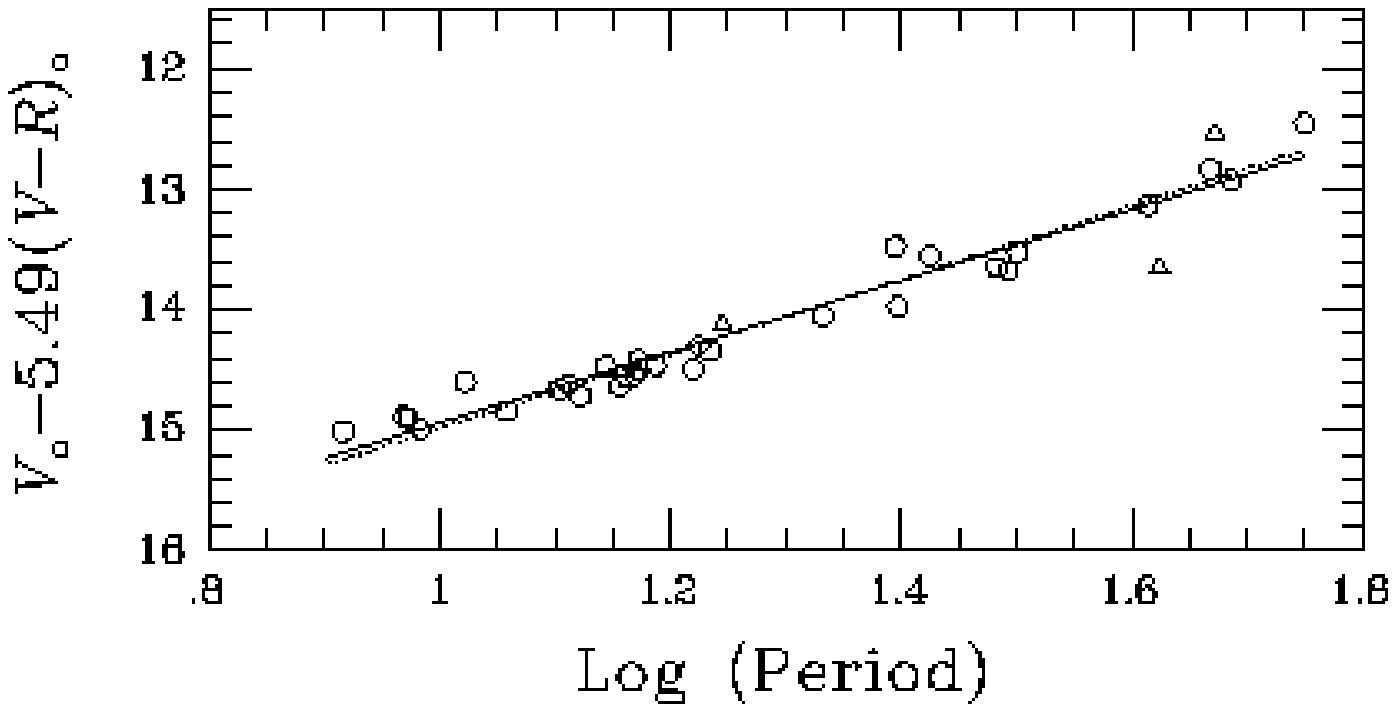}
\end{minipage}
\end{figure*}

\clearpage

\end{document}